%% file: paper.tex
\documentclass[letterpaper,twocolumn,10pt]{article}
\usepackage{usenix2019_v3}

\usepackage{caption}
\usepackage{upgreek}
\usepackage{listings}
\usepackage{amsmath}

\usepackage{color}
\usepackage{comment}
\usepackage{ulem}
\usepackage{graphicx} %use graph format
\usepackage{subfigure}
\usepackage[ruled,vlined,linesnumbered]{algorithm2e}
\usepackage{multirow}
\usepackage{makecell}
\usepackage{setspace}
\usepackage{lipsum}
\usepackage{titlesec}
\usepackage{xcolor}
\usepackage{xspace}
\usepackage{textcomp}

\usepackage{pifont}
\definecolor{mybg}{rgb}{0.82,0.91,0.84}
\input{defs}

\begin{document}
%-------------------------------------------------------------------------------

%don't want date printed
\date{}

\title{
\Large \bf 
QStack: Re-architecting User-space Network Stack to Optimize CPU Efficiency and Service Quality
}

\author{
{\rm Wenli Zhang}\\
Institute of Computing Technology, \\Chinese Academy of Sciences (ICT, CAS)
\and
{\rm Yifan Shen}\\
ICT, CAS\\
University of Chinese Academy of Sciences (UCAS)
\and
{\rm Hui Song}\\
ICT, CAS
\and
{\rm Zhao Zhang}\\
ICT, CAS
\and
{\rm Ke Liu}\\
ICT, CAS
\and
{\rm Qun Huang}\\
Peking University
\and
{\rm Mingyu Chen}\\
ICT, CAS\\
UCAS
} % end author

\maketitle
% \input{subsections/abstract.tex}
%-------------------------------------------------------------------------------\begin{abstract}
    TCP/IP network stack is irreplaceable for Web services in datacenter front-end servers, and the demand for which is growing rapidly for emerging high concurrency network service applications (including Internet, Internet of Things, mobile Internet, etc.) especially. The existing network stack schemes often face the dilemma between the data center server resource utilization (i.e., high CPU efficiency) and application service quality (i.e., low tail latency). We break this dilemma via a flexible architectural design QStack, which simultaneously achieves CPU efficiency and low tail latency in user-space network stack for front-end datacenter server. QStack proposes elastic framework and application definable full-datapath priority, such that the network stack collaboration among CPU cores horizontally and coordination across network layers in fine grained vertically on demand. We prototype QStack on commodity servers. Testbed experiments demonstrate the effectiveness of QStack over state-of-the-art user-space network stack designs.
% \end{abstract}

%-------------------------------------------------------------------------------
%\section{Introduction}
%-------------------------------------------------------------------------------
% \input{subsections/introduction.tex}
%-------------------------------------------------------------------------------
\section{Introduction}
\label{sec:introduction}
%-------------------------------------------------------------------------------

With the development of critical technologies in the 
Internet of things (IoT), 
more new applications are invented and becoming widely used 
(\textit{e.g.}, camera surveillance, digital healthcare, smart grid, smart city)~\cite{iot-2016}.
The number of clients (\textit{e.g.}, IoT devices) connected to clouds 
increase dramatically, \textit{e.g.}, from $20$ billion in 2017 to $75$ 
billion in 2025~\cite{WEB:idc-iot}. 
The ever-increasing client population and ever-evolving service portfolio 
lead to the growing intensity, concurrency, 
and diversity of the workload destined to datacenters.
Therefore, 
the front-end server -- the gateway of datacenters~\cite{frontend-imc-11}
is becoming more and more important 
for datacenter operations, which is required to process such workload
efficiently.

However, existing network stacks prevent front-end servers from performing efficiently under such circumstances. The conventional wisdom is that the kernel-based network stack is notorious for its large overhead~\cite{arrakis-2014}, 
and is not suitable for workloads with high concurrency~\cite{mtcp-2014}. Robert Graham pointed out when he proposed the C10m (10 million concurrent connections) problem~\cite{2013c10m} that, "The secret to 10 million concurrent connections  – the kernel is the problem, not the solution", and the kernel problem involving packet handling, memory management and processor scheduling. Recent user-space network stacks were shown to improve 
tail latency and CPU efficiency substantially under various workloads. However, as we demonstrate later, designs of existing user-space network stacks still have two deficiencies:

First, existing network stacks are usually not qualified for independent scheduling, and associated with application scheduling while adjust the CPU resources allocation according to application characteristics. The scheduler either dedicates fixed CPU cores for network protocol processing or binds network and application logics as a whole unit.
Given the various incoming network traffic and application workload in datacenter, such static allocation may result in either CPU inefficiency or bad tail latency: Shinjuku, Perséphone, and Shenango~\cite{shinjuku-2019,idle-sosp-2021,shenango-2019} 
run their network stacks on one or several dedicated CPU cores. 
If the incoming network workload is mild, the CPU resource is wasted due to 
busy-polling. If the network workload is heavy, 
the current CPU cores might be not sufficient to process the peak workload with 
desirable latency. mTCP, IX, ZygOS, and F-Stack~\cite{mtcp-2014,ix-2014,zygos-2017,WEB:F-stack} dedicates network stack and application computation in static coupled mode on each CPU core. Although this design improves data locality~\cite{ix-2014}, it  results in either CPU inefficiency 
if the application has few workloads to process, 
or increased tail latency and even packet drops at NIC if 
the application logic occupies more CPU cycles while the 
network stack is waiting for critical processing. 

Second, existing network stacks process the network requests equally, first come first served. 
Although this manner could improve CPU inefficiency and tail latency to some extent, 
it is not optimal when the service time of each request varies significantly. One common problem is Head-of-Line blocking (HoL), that is, a critical and short-running request may be blocked by a previous long-running request.
Although this can be partially resolved 
by adopting like preemption~\cite{shinjuku-2019} and 
task-stealing~\cite{zygos-2017, shenango-2019}, 
it would only be triggered 
when the application performance has already been deteriorated, 
and results in more overhead as a side-effect 
~\cite{shinjuku-2019, zygos-2017, shenango-2019}. 

In summary, there are two issues both related to CPU resource allocation: the CPU resource allocated to network stack is static, the resource allocated for each request is uncorrelated with the application semantic. 
To address these two problems, we propose QStack, 
an elastic user-space network stack with application definable priority support, which features two novel designs:

\begin{itemize}
\item{\textbf{Elastic framework.}}
QStack designs elastic framework to support dynamic resource allocation from as low as one core to the whole server, which improves CPU efficiency in fluctuant datacenter workload. Fastcalldown ensures network stack and application sharing in one core effectively with no packet drop to get both the timeliness of interrupt and the low overhead of polling. It is suitable for the limited core number scenario like virtualization especially. Dynamic detect would be enable QStack scaling to the whole server, in which we get 10 million concurrent connections work well in one server.

\item{\textbf{Application definable full-datapath priority.}}
QStack designs application definable full-datapath priority to dynamically customize the request feature labels to guide the prioritization of some requests/responses in low overhead, which guarantees the QoS of delay-sensitive requests (e.g. low tail latency). As well, we also combine priority strategy with elastic framework as extension. 

\end{itemize}

We prototype QStack as a user-space TCP/IP network stack over DPDK~\cite{WEB:dpdk}, and port three applications to QStack including IoT front-end Server, HTTPs service, and the popular Web server Nginx. 
\summary{brief evaluation}
The experimental results justify that the effectiveness of QStack over 4 state-of-the-art user space network stack designs. For example, QStack is the only user-space network stack that can dynamically adjust CPU resources according to load characteristics while improve CPU utilization and service tail latency simultaneously (See in Figure~\ref{fig:evaluation_elasticframework}). 
And QStack is able to support over 10 million concurrency, using a cloud-supported IoT application scenario, with an order of magnitude improvement over user-space network stack mTCP and X86 Linux kernel TCP stack (See in Figure~\ref{fig:evaluation-iotepserver}) in a mainstream commercial server taking 50ms as the tail latency threshold.

\def\widtha{0.20\textwidth}
\def\widthb{0.25\textwidth}
\def\widthc{0.21\textwidth}
\begin{table*}[]
  \small 
  \caption{Summary of design decisions in QStack}   
    \begin{tabular}[c]{p{0.15\textwidth}p{0.07\textwidth}p{\widtha}<{\centering}p{\widthb}<{\centering}p{\widthc}<{\centering}}
    \hline
    design decision & rationale/benefits & challenges & solution & overhead \\
    \hline
    \textbf{Elastic framework}, support dynamic resource allocation from as low as one core to the whole server &

        Improve CPU efficiency in fluctuant datacenter workload &

        \begin{tabular}[t]{p {\widtha}}1) How to guarantee the  critical subtask timeliness without NIC interruption or time slices, while avoiding using dedicated polling cores? \\ 2) How to support resource reallocation according to the real-time load?\end{tabular} &

        \begin{tabular}[t]{p {\widthb}}1) Fastcalldown: guarantee the critical subtask timeliness under the application logic of uncertain duration with low overhead\\ 2) Dynamic detect: dynamic detect workload on both network stack and application to reallocate resource\end{tabular} &

        \begin{tabular}[t]{p {\widthc}}1) Fastcalldown: 22ns mostly, << 2.2\% extra overhead, in finer granularity and reducing unnecessary operations than coroutine switch \\ 2) Dynamic detect: reallocate CPU cores within 2 statistic periods\end{tabular} \\
    \hline
      \textbf{Application definable full-datapath priority}, dynamically customize the request feature labels to guide the prioritization of some requests/responses in low overhead &

        Guarantee the QoS of delay-sensitive requests (e.g. low tail latency) &

        \begin{tabular}[t]{p {\widtha}}1) How to efficiently provide application-level characteristics as the basis for priority scheduling across multiple layers in the network stack?\\ 2) How to schedule requests with priority feature across layers in the network stack?\end{tabular} &

        \begin{tabular}[t]{p {\widthb}}1) Fastcallup: extract priority feature according to application customization dynamically, collaborate with elastic framework\\ 2) Full-datapath priority: prioritization based on request feature labels across all network stack layers to avoid queueing delay blocking\end{tabular} &

        \begin{tabular}[t]{p {\widthc}}1) Fastcallup: around several nanoseconds (e.g. read certain flag bits) to hundreds of nanoseconds (e.g. simple keyword matching) usually\\ 2) Full-datapath priority: effective priority scheduling according to the characteristics of each layer\end{tabular}\\
    \hline 
    \end{tabular}
    \label{tab:design-decision}
\end{table*}

%-------------------------------------------------------------------------------
%\section{BackGround \& Motivation}
%-------------------------------------------------------------------------------
% \input{subsections/background.tex}
\section{Background and Motivation}
\label{sec:background}
%%%%%%%%%%%%%%%%%%%%%%%%%%%%%%%%%%%%%%%%%%%%%%%%%%%%%%%%%%%%%%%%%%%%%%%%%%%%%%%%
\subsection{System Model}
\label{sec:model}
We focus on the network stack design of a host server that is required to 
process fluctuant network traffic with high concurrency and  instantaneous burst. 
A typical example is the front-end server -- the gateway of datacenters, which needs to process and 
forward end-user requests of various types of workloads destined to the back-end servers of datacenters~\cite{frontend-imc-11}.

For a typical datacenter architecture, 
a front-end server is deployed to provide entry service for every request, 
such as filtering, parsing, decomposition~\cite{microservice-asplos-2019}.
For example, 
\texttt{Router}~\cite{usuite-2018} parses and filters query payload into 
different query types, distributes keys uniformly across backend servers, and then routes the requests. In the Facebook social network, 
front-end servers provide Web services, load-balancing, and caching to users
~\cite{DBLP:conf/sigcomm/RoyZBPS15}. If the requests are not met in front-end servers, they will be forwarded to back-end clusters.

%-------------------------------------------------------------------------------
\subsection{Workload Characteristics}

\textbf{Fluctuant workload.}
As measured in ~\cite{traffic-in-the-wild-IMC-2010}, 
the link utilization is subject to time-of-day and day-of-week effects across 
different kinds of data centers. 
Tencent also reported that the amount of requests in unit time during peak hours is about 3 
times larger than the daily average~\cite{microservice-overload-wechat-socc-2018}.
Therefore, it is necessary to consider dynamically adjust the CPU resource for the network stack according to the load.

\textbf{Various request types.}
The requests destined to frontend servers have different service times and 
are associated with different SLOs (Service Level Objective). 
One category of requests are finished in a short period, 
\textit{e.g.}, frontend servers simply parse the user queries and forward them to backend servers with load-balance, 
which takes only a few microseconds~\cite{frontend-imc-11}.
There also exists one category of requests that are CPU-intensive, 
which take a long time to complete, 
\textit{e.g.}, in HDSearch~\cite{usuite-2018}, 
frontend servers might extract a feature vector for the query
image with TensorFlow~\cite{tensorflow-osdi-2016}.  
In addition to serving time, 
different requests have different SLOs, 
\textit{e.g.}, the users are more sensitive to the latency variation of requests 
with smaller basic latency~\cite{DBLP:conf/sigcomm/Zhang0KGJ19}; 
and in some applications like IoT service, 
a large amount of requests like state update and keep-alive packets are not 
latency-sensitive, 
while a small fraction of interactive requests is latency-critical\cite{lns-jcst-2020}.
Therefore, it would be better for the network stack design to schedule the tasks in fine granularity to avoid high tail latency.

\textbf{High cost of packet drop.}
The end-users from external networks like the Wide Area Network (WAN) connect to frontend servers, 
using long-lived TCP connections.
The end-to-end delay is large, thus packet drop is not desirable as 
retransmission substantially increases the response time.
Therefore, a good network stack for frontend server should avoid endpoint packet drops.

\subsection{Opportunities and Motivation}
\textbf{Dynamic CPU Adjustment for Network Stack.}
\label{sec:bg-architecture} Processing front-end services with high concurrent load requires sufficient CPU resources to ensure timely processing of the network stack. Backend low latency services need to put the application processing  and the network stack processing of the same request on the same core to improve Cache locality. Batch applications need to allocate sufficient computing resources for the upper application processing and place them on a separate core to avoid interference with the network stack and other application processing. In addition, the fluctuation of data center load leads to the change of CPU resource requirements for network stack processing and application processing over time.The network stack needs to be able to adjust the CPU allocation dynamically. 
Motivated by the CPU allocation problem in existing work, it would be better to design network stack to be elastic, \textit{i.e.}, dynamic CPU adjustment for network stack processing. Ideally, the CPU resource for network stacks and applications should be allocated independently and dynamically according to the network and application workload, respectively. 

\textbf{Extract application feature for scheduling.}
\label{sec:bg-schedule}
Workload characteristics are usually most abundant and intuitive at the application level, and the global view of the network stack would be the best point to extract features. The network stack would be a sweet spot to do extraction mechanisms (in \S~\ref{sec:bg-relatedwork}) 
to obtain the application intrinsic features, 
since the network stack process all the packets and maintain status for all the connections, which has a global view with abundant information, \textit{e.g.}, the congestion status, the packet length, the sequence number, \textit{etc.} 
Most of all, all the requests would be processed by the network stack prior to scheduling, to obtain sufficient information successfully. Therefore, it would be ideal if we could define the load characteristics from the application perspective, and then complete the feature recognition in the network stack processing window as soon as possible.

These considerations led to particular design decisions for QStack,
which we summarize in Table~\ref{tab:design-decision}.

%-------------------------------------------------------------------------------
%\section{Qstack Design}
%-------------------------------------------------------------------------------
% \input{subsections/design.tex}
\section{QStack Design}
\label{sec:design}

We state our design goals and assumptions (\S~\ref{sec:designoverview}) , and show how elastic framework (\S~\ref{sec:design-elastic_architecture}) and application definable full-data-path priority (\S~\ref{fullpathpriority}) to achieve both resource efficiency and service quality respectively.

\begin{figure*} [htbp]
    \centering 
    \includegraphics[width=15cm]{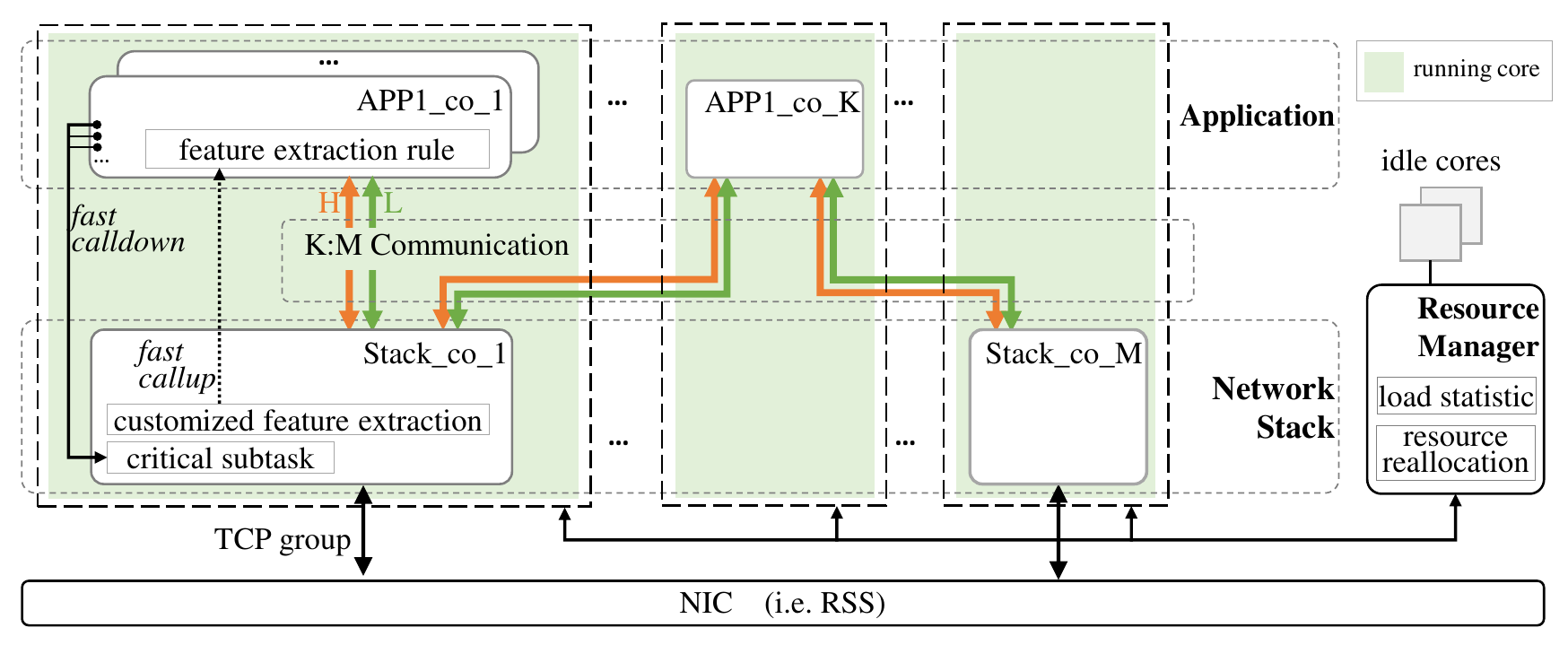}
    \vspace{-0.3cm}
    \caption{QStack Architecture} 
    \label{fig:design-overview}
    % \vspace{-0.4cm}
\end{figure*}

\subsection{Design Overview} 
\label{sec:designoverview}
As depicted in Fig 1, QStack is a flexible user-space network stack architecture that can flexibly adapt to the application characteristics, mainly including multiple layers such as driver, TCP/IP network stack, event framework, etc. It is suitable for undertaking the diversified service needs of front-end servers in the data center. It can effectively take into account the CPU resource utilization and service tail latency from an architectural perspective, by re-architecting the collaboration between the network stack and the upper application. It supports to allocate CPU resources for the network stack and the upper application respectively, so as to meet the architecture requirements of multiple data center applications running on the same user-space network stack simultaneously. At present, it supports multi-threaded application mode, and provides BSDsocket-like data plane interface and epoll-like event framework interface. The names of these interfaces are similar to the native interfaces, but "q\_" is added as prefix. The logic and parameters used shall also be consistent with BSD and epoll interfaces as far as possible, to facilitate developers to migrate existing applications or develop new applications. Actually, epoll is the most commonly used event framework on IO multiplexing programming model in Linux by data center applications, and socket interface is the most common data receiving and sending method in data center applications. In particular, QStack also provides an interface q\_get\_Event() to receive a single event, which ensures the trigger frequency of fast callback by increasing call frequency, and provides fine-grained event framework priority scheduling.

\textbf{Assumptions.} We make several design assumptions.

\begin{itemize}
\item We focus on front-end applications that rely on traditional TCP, and scenarios applicable to RDMA for internal use in data center out of scope for this paper.
\item We focus on network stack and front-end application processing, and does not involve too much hardware and back-end applications.
\item We only use one CPU for the microbenchmarks to exclude the impact of NUMA architecture. 

\end{itemize}

\subsection{Elastic Framework}
\label{sec:design-elastic_architecture}

QStack designs the elastic user-space network stack architecture to support dynamic adjustment of computing resource allocation for the network stack and application on demand, to adapt to the fluctuation of the data center workload. It avoids overloading due to insufficient computing resources during peak load periods, and avoids busy polling of empty queues during low load periods, which will give consideration to both CPU resource utilization and quality of service. Notably, If the application is able to share one core with the netowrk stack, it would be especially meaningful for scenarios with limited cores, such as the virtualization scenarios widely used today.

 However, there are two practical challenges:

\begin{itemize}
\item 1) How to avoid application blocking within a shared core? 
When sharing one core between network stack and application, once the application takes a long period for processing, it will block the critical processing of network stack, such as NIC access and TCP timer checking, and result in packet drop, which will be more serious with higher retransmission delay cost for WAN links. As a contrast, in the kernel network stack, the NIC hard interrupt is used to collect data packets and the soft interrupt is used to process data packets, which ensures the real-time performance of the protocol critical processing, but brings high processing overhead. Or in the user-space network stack, some fixed dedicated processing cores are used for busy polling to get low overhead.

\item 2) How to detect and reallocate resources according to the real-time load, and how to smoothly adjust the collaboration between network stack, NIC queues, and application logics? As well, the system should keep the consistency of the interfaces and data structures during the transition and avoid locks and blocking for efficiency at the same time. 
\end{itemize}

\textbf{Fastcalldown.}
QStack designs fastcalldown, a cross coroutine (thread) function call processing mechanism within the core, to deal with the problem that how to guarantee the timely processing of urgent subtasks. If the application processing time is long enough, the NIC queue would overflow and packet drop would occur.

\begin{figure}[t]
    \centering 
    \includegraphics[width=6cm]{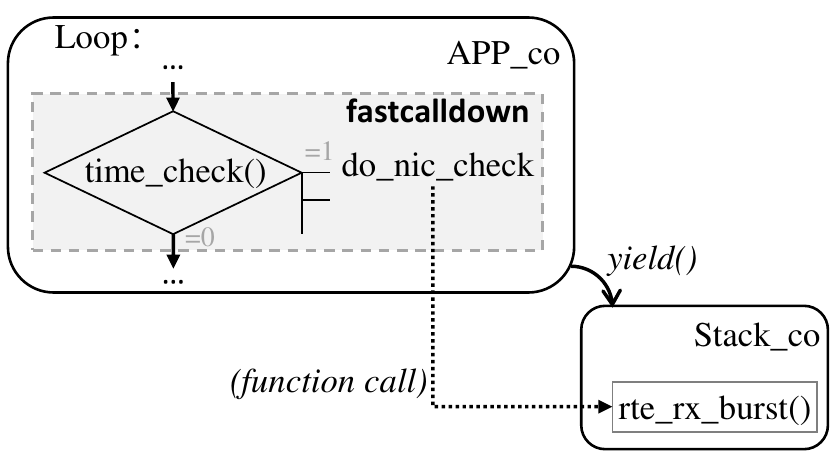}
    %\vspace{0.3cm}
    \caption{A fastcalldown example.} 
     
    \label{fig:Fastcalldown}
    %\vspace{0.6cm}
\end{figure}

As shown in Figure~\ref{fig:Fastcalldown}, a typical scenario is how to receive packets timely and avoid the NIC queue overflow without interruption during application processing. When a fastcalldown is triggered, it first obtains the current system time and checks the time interval is whether within the threshold. If NIC check interval is timed out, fastcalldown would directly call the packet receiving function to collect data packets to the driver buffer for temporary storage. If the TCP process interval is timed out, fastcalldown would directly process a batch of packets (for example, 64 packets), process a batch of application requests (for example, establish a new connection or close a connection), and then send a batch of packets. Fastcalldown also checks the running time of the current application coroutine. If it exceeds a certain threshold (for example, 10 milliseconds), it would trigger the rescheduling to yield the current processing core to other coroutines/threads.

Fastcalldown mainly uses in two ways: 1) Implicit call, which implicitly inserts the function call of fastcalldown into the system function interface (such as epoll\_wait(), recv(), send(), etc.) called by the application, which transparent to the application; 2) Explicit call: when the application processing time is obviously uncertain, for example, in a long loop of application, the developer should explicitly insert fastcalldown at a certain frequency for critical subtask checking. Of course, if the developer clearly knows that an application will have a large section of code that cannot insert fastcalldown into, it is recommended to arrange it to a separate processing core based on elastic framework to avoid blocking other network stack and application processing. In the future, we hope to identify appropriate points and automatically insert fastcalldown in compiler, so that data center applications can be more easily transplanted or developed on QStack.

Take implicit call as an example. Generally, the execution time of a request is from a few microseconds to tens of microseconds. Each request will trigger fastcalldown when acquiring events, receiving packets, and sending packets. On average, the interval of fastcalldown is about microseconds and 10 microseconds.

Mostly, fastcalldown just get the current time, check the timestamp and then return to continue before the scheduling is needed, and the cost is about 22 nanoseconds. Taking the Intel NIC 82599 as an example, it takes about 100ns to check an empty queue, and about 1us to process 32 packets in a batch; In extreme cases, it takes more than 200 microseconds for 10G line speed of 64B packet to fill one NIC queue with a length of 4096. While QStack often uses multiple queues to receive packets, so it would absolutely ensure no packet drop taking 200 microseconds as the packet receiving threshold. Suppose that the fastcalldown insertion is performed at microsecond interval, the extra overhead would be about 2.2\%, which will be much lower in the actual scenario.

Compare the overhead with the coroutine mode in Demikernel ~\cite{2021The}. Taking 200 ns as a period, assuming that N requests can be processed during this period, the Demikernel has a 12 cycles (6 ns) yield cost after each request and a call to rte\_rx\_burst() receiving packets. Let’s take that cost as empty NIC queue check, about 100ns. Then total NIC check cost would be (6+100)$\times$N=106N ns. For fastcalldown, The implicit call would be triggered at three places including collect events, q\_recv() and q\_send(). Usually, only timestamp check is performed each time and the cost is 22ns. It is assumed that rte\_rx\_Burst() is performed at the last time to do NIC check and also use the cost 100ns. Then the total cost would be 22$\times$3$\times$N+100=100+66N ns. Then, as long as N>2.5, that is, when receiving more than three requests, the cost of the fastcalldown mode will be better than that of the coroutine mode in Demikernel. For this reason, fastcalldown would be more adaptable to the application duration and would be more adapt to the real data center applications. While, some other work would assumed the application processing in a short running time, for example, the Demikernel application coroutine is assumed in microseconds. 

In summary, fastcalldown uses the embedded calldown technology to ensure the real-time performance of critical processing in network stack and coroutine scheduling with low overhead, while taking into account the timeliness of interrupt and time slice in OS and the low overhead of user-mode polling, but avoiding their high overhead or wasting CPU resources. Fastcalldown would be applicable to all scenarios where "coroutines/threads  are on the same CPU core and in the same namespace, while one of them has a critical subtask to process  timely".

\textbf{Dynamic detect.} Oriented to fluctuant workload in data center, QStack uses resource manager (running on the core without application and network stack) to monitor the real-time load of network stacks and applications, and dynamically formulate CPU resource allocation strategies, dynamically adjust the number of application and network stack coroutines, and manage the number of CPU cores used by QStack. It would avoid overloading due to insufficient computing resources during peak load, and avoid busy polling empty queues to waste CPU during low load. In order to support the dynamic resource ajustments of application and network stack, for the upper, QStack uses the K: M parallel and lock free communication framework based on MPSC (Multi Production Single Consumption)  queues to achieve parallel, dynamic and efficient communication and task distribution among applications and network stacks; and for the lower, QStack uses TCP flow groups got by Intel RSS hash to achieve low overhead dynamic correlation among network stacks and NIC queues. And the smart NIC is optional to achieve more accurate, fine-grained and low-cost dynamic stream group mapping.

QStack includes a simple default scheduler, which can suspend, wake up or migrate coroutines/threads to the specified core, and poll all application coroutines in the form of round robin for scheduling. The scheduler can obtain the number of backlog events in the event queue corresponding to the current application coroutine through the specified interface, and count up the request execution time at the application layer; obtain the data packets sent and received by the network stack, buffer backlog, etc., and determine whether there is overload or resource waste. If a CPU core has been overloaded for a period of time (for example, two consecutive scheduling periods), the application load on the processing core needs to migrate partly to other processing core. The current backlog of events will be processed in local. If the current application coroutine switches the entire processing core to the suspended state after processing all the remaining events, the CPU will be scheduled by the kernel and used by other applications on the server. The scheduler can also dynamically create/destroy coroutines/threads. However, due to the high overhead, it is generally recommended to apply for enough when the system starts, and change the number of coroutines/threads in the way of suspending/waking up. In practical applications, it is recommended to customize a scheduler according to the requirements, in which the monitored load characteristics and resource adjustment strategies all can be customized on demand.

In the process of dynamic scheduling and allocation of CPU resources, the main difficulty is how to efficiently detect and adjust the mapping relationship dynamically among the network stack coroutines and the NIC queues, and the upper application coroutines.

\subsection{Application definable full-datapath priority}
\label{fullpathpriority} 
QStack designs an application-defined scheduling feature extraction mechanism, which supports upper-level applications to flexibly customize analysis functions for feature extraction according to application and load characteristics. The application-defined function would be registered in the network stack, and the feature extraction points could be set at the NIC, driver, TCP/IP and event framework layer to get scheduling feature extraction as early as possible and get prioritization. Usually, upper layer would get more complicated information, for example the connection related information should be extracted in TCP layer not in NIC or driver layer. This extraction mechanism would provide rich status information involving data packets and TCP connections in real time for scheduling, which enables flexible scheduling feature extraction for requests in different load scenarios with no need to customize the network stack. 

At the same time, the scheduling feature extracted from the network stack would be transferred in the form of a feature label along with the request in the full data path. For stages that may cause queuing congestion, such as receiving and sending buffer in driver layer, event queue and thread scheduling, priority queues are provided so as to ensure that once delay sensitive requests are identified, all processing stages after the extraction point could be prioritized. 

On this basis to extend, combined the above elastic framework with the request feature label, priority diffluence, priority check based fastcalldown and other extensions are added, which will further avoiding the delay sensitive requests from being affected by the queuing congestion of hybrid workloads.

However, it has to face the following two challenges:

1) How to flexibly transfer the scheduling semantics of the application layer to the network stack layer, so that application can accurately identify the scheduling type of different requests according to customized scheduling feature analysis strategy? It is necessary to extract the abundant information from the network stack with an efficient abstraction interface.

2) How to identify and label the request feature as early as possible, and schedule the corresponding request processing at all processing stages especially where may be blocked? Queuing and blocking interference caused by hybrid request workload may occur at any stage of request processing, including driver receiving and sending packets, packet processing, thread scheduling, etc.

\textbf{Fastcallup.} QStack designs fastcallup, a function callback that identifies the request features in the network stack defined by the application. The network stack calls the callback function defined by the application in driver, TCP processing and other stages to identify the characteristics of the packets and write them as feature labels with the packets, and then performs priority scheduling in the request packet processing, event scheduling, thread scheduling and other stages according to the identified priority. It would avoid latency critical requests from queuing and blocking delays caused by insensitive requests. The lower the level, the fewer features could be used, but it could reduce the queuing delay more.

\begin{figure}[t]
    \centering 
    \includegraphics[width=7cm]{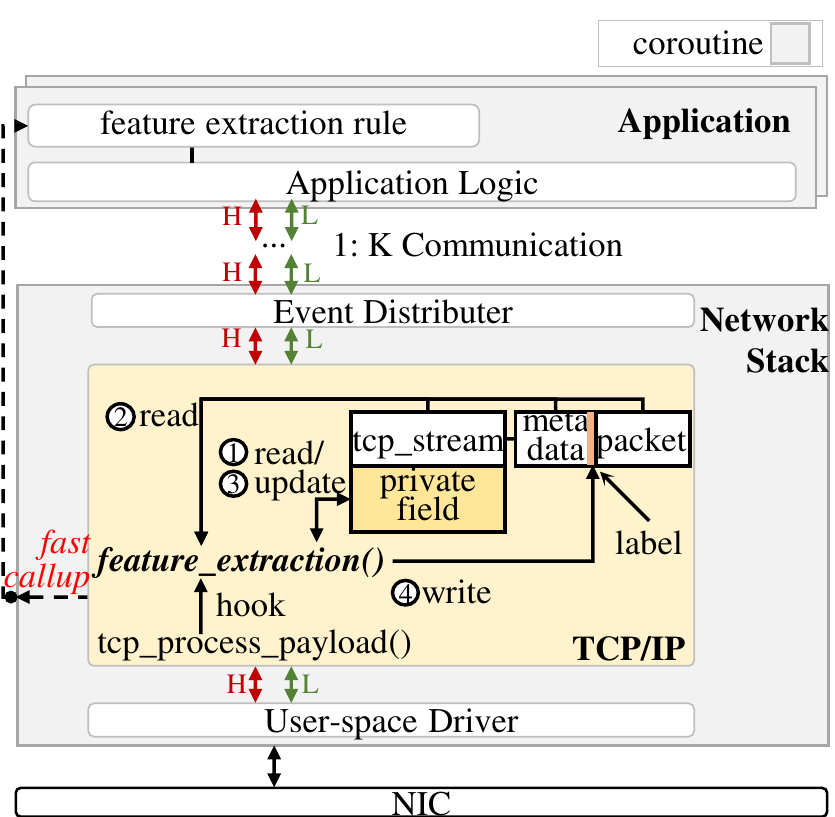}
    %\vspace{0.3cm}
    \caption{A fastcallup example in TCP/IP layer.} 
    (An example on message boundary identification.\ding{172}read the message boundary from private field written by the last packet in the same message,
    \ding{173}read packet data,
    \ding{174}update private field,
    \ding{175}write the priority returned by \textit{tcp\_extraction()} to the packet metadata
    )

    \label{fig:fastcallup}
    %\vspace{0.6cm}
\end{figure}

Taking the TCP/IP layer method as an example, it is located on the critical path of each request processing and provides connection-oriented flow support, reliability, flow control, multiplexing and other services for applications. Compared with the driver layer which only get packet level information, it can also obtain stream level state and semantic information, and can communicate directly with the upper application through socket abstraction. As shown in Figure ~\ref{fig:fastcallup}, the application register the scheduling feature extraction function for each TCP connection to the user-space network stack in the form of a registration callback function tcp\_extraction(). And the new scheduling feature extraction function could be updated for the current TCP flow according to the status change. Whenever the TCP/IP network stack processes packets with payload, it will analyze the characteristics according to the callback function. The analysis result is written into the metadata of the packet in the form of label, and transferred to the events added to the event framework and to be scheduled according to the label. Besides, we propose the design of priority identification private field for each TCP stream to store the intermediate state of priority feature analysis (for example, priority identification for a single message across packets). The private field length can be up to 64 bytes (a Cache line), which can be read and written by the fastcallup function in TCP layer, and support the upper layer applications to access through the socket interface.

It looks a bit like eBPF at first glance. In fact, eBPF only takes effect in  specific locations of the OS kernel, and is suitable for such as packet filtering, distribution and tracking, etc., and the results are uniformly accessed in specified space. While QStack works in the user mode as a whole. A hook (function pointer) for the callback function is reserved in the TCP stream structure to facilitate fastcallup to set features in TCP connection. The extracted feature is in the form of lightweight label, and is transmitted with the request along the full path across layers to guide the priority scheduling.

\textbf{Full-datapath priority support.} Priority feature can be extracted at all levels of the server network stack, and can also be transferred to use or updated in subsequent layers, so as to avoid the impact of queuing or blocking delay on tail latency of high priority requests. 1) Driver layer priority queue support. The driver provides buffers with different priorities in both send and receive, and high priority packets can be processed first. After the packets are prioritized, the response latency of specific packets can be significantly reduced. 2) TCP out of order priority support. QStack gets packets from the high priority NIC queue preferentially. In order to avoid the impact of HoL blocking caused by TCP sequencing, QStack puts high priority TCP packets into a separate receiving buffer and provides a socket interface with priority for out of order reception, so that the upper layer application can directly receive high priority request packets, thus avoiding the queuing delay caused by blocking. 3) Event framework priority support. QStack event framework provides priority support. When the network stack sends a new event to the upper layer application, it can write the event priority. The high priority events will be received by the upper layer application first, to reduce the queuing delay of priority requests.

As extension, it supports 1) priority diffluence, in order to avoid blocking delay caused by complex requests occupying CPU for delay sensitive requests running on the same core, QStack designs feature label diffluence on the basis of elastic framework. A new interface q\_epoll\_ctrl() with a priority field is added, which specifies to receive high or low priority events when binding sockets. This benefit from the K: M communication among network stacks and applications in elastic framework. 2) Priority based fastcalldown. If multiple applications are running on the same processing core and different scheduling types of requests are processed at the same time, suppose that new delay sensitive request arrives at the server during the processing of low priority request, since the low priority request is occupying the CPU, the upper layer applications cannot timely collect delay sensitive requests for processing, which would cause queuing delay. On the basis of fastcalldown in elastic framework, QStack adds coroutine priority check. During the execution of fastcalldown, if there is higher events coming than the current, it would yield () immediately to the higher scheduling priority for processing. Here, the request feature label is required to provide scheduling feature. Fastcalldown is required to perceive higher priority requests in the same core in a timely manner during task execution. As well, it also requires the coroutine with stack to switch fast.

%-------------------------------------------------------------------------------
%\section{Qstack Implementation}
%-------------------------------------------------------------------------------
% \input{subsections/implementation.tex}
\section{Implementation}
\label{sec:Implementation}

We implement a prototype of QStack. QStack divides request processing into upper application processing and network stack processing separately as coroutine. On each core, one (Linux/kernel level) thread is set affinity to support running 0 or 1 network stack coroutine and 0 or more application coroutines. Superior to the static full data path priority of LNS \cite{lns-jcst-2020}, QStack implements the application definable priority involving the driver and TCP/IP layers, and event framework as an optional extension. The resource manager is optional to adjust resources more intelligently. It is worth mentioning that, QStack provides full-datapath zero copy and full-stack lock free, i.e MPSC queue, till application layer including sending and receiving to reduce the cost of memory copy and resource contention.

\textbf{Network stack processing}. QStack implements user-space network stack as key function considering both low overhead and high efficient. We leverage DPDK to remove kernel-space overhead and optimize multi queue to support. The fastcalldown function is implemented in combination with the pre-allocated large memory pool, which is enough to sustain short-term burst traffic with low overhead in software, so as to avoid packet drop in NIC. We use about 1000 lines of mTCP code, mainly including TCP protocol header parsing and state machine processing to save time. 

% \textbf{K:M communication}. We optimized epoll to support K:M communication between protocol stack and application for elastic framework without lock and atomic. The multi-production-single-consumption (MPSC) queue optimized to lock free supports M protocol stack threads to send events to the same application thread; Simultaneously, the combination of multiple MPSC queues supports 1 protocol stack thread to send events to K application threads. In the K:M communication mode, the principles of locality and load balancing on event distribution and scheduling are taken into account, to achieve efficient communication between protocol stack threads and application threads in the elastic architecture.

\textbf{Application development and porting}. 
To demonstrate the ease-of-use and flexibility of QStack, we developed two applications on top of QStack including IoTServer and HTTP\_TLS, and ported one existing application, the popular Web server Nginx.

IoT frontend server. It works in the form of a typical IoT frontend server, and has similar logic with Keyvalue Router. The implementation of IoTServer takes about 1100 LoCs. Its primary functionality is to route client queries, to suitable key-value store servers in backend Redis clusters, and send back response. IoTServer receives the users’ query payload with QStack’s socket API and get the priority feature with event API, which is provided by the registered analysis function. 

HTTP\_TLS with encryption and decryption. It is a typical network function HTTP\_TLS, for HTTPS messages encryption/decryption with OpenSSL, a library for SSL/TLS. QStack leverages the scheduling feature extraction in network layer, which decode only the header of message at network layer to identify the priority, so that avoid the queuing delay caused by heavy overhead from the decryption/encryption of low-priority requests.

Web server Nginx. QStack ports a real application, Nginx, an open-sourced event-based Web server. We modified Nginx from multi-process-based to an equivalent way single-process-based with multithreading to adapt current QStack. We replaced the Nginx Socket and event APIs with QStack's counterparts, which involves roughly 114 LoCs.

\textbf{Resource Manager}. It is optional for more efficient or intelligent resource utilization since the local core could be responsible for coroutine state and slow tentative increase and decrease the number of coroutines. QStack developes a default resource manager as an independent thread to get load characteristics and resource utilization with intelligent regulation from a global perspective and to feedback decision quickly and accurately. 
QStack implements K:M communication with MPSC queues among network stacks and applications to support flexible resource adjustment. QStack leverages TCP group got by RSS hashing to support fast flow migration among network stacks and NIC queues by changing the pointer. However, it is limited by the number of NIC queues (for example, Intel 82599 only supports 16 queues), the granularity of the CPU resource management for the network stack is relatively coarse, smart NIC would be a better choice later.

%-------------------------------------------------------------------------------
%\section{Evaluation}
%-------------------------------------------------------------------------------
% \input{subsections/evaluation.tex}
\section{Evaluation}
\label{sec:evaluation}

We conduct testbed experiments that compare QStack with 4 state-of-the-art user space network designs (including mTCP, IX, Shenango and F-Stack) and Linux kernel TCP. The experiments involve three aspects: 1) elasticity (Exp\#1-Exp\#2), 2) priority (Exp\#3-Exp\#6) and 3) applications (Exp\#7-Exp\#9). We summarize our findings for QStack:
\begin{itemize}
\item It achieves elastic scalability efficiently on allocating resources independently for network stack and applications on demand (Exp\#1). It supports the full-datapath priority of the server network, and the earlier the priority is identified, the more obvious the advantage is (Exp\#5). And it can effectively support cross packet priority (Exp\#3). 
\item It can avoid packet drop better based on fastcalldown design than mTCP and IX (Exp\#2), and improve CPU efficiency.
\item It not only guarantees the response tail latency of delay sensitive requests, but also improves the CPU utilization efficiency compared with Shenango (Exp\#4), based on fastcallup design to support application definable full-datapath priority. Priority diffluence shows the special advantages of the fastcallup to collaborate with elastic framework (Exp\#6). And it has obvious advantages in supporting high concurrency and low tail latency, which is able to support over 10 million concurrency in a mainstream commercial server taking 50ms as the tail latency threshold in a typical IoT operation scenario, an order of magnitude better than user space network stack mTCP, and X86 Linux kernel TCP stack (Exp\#7).
\item It supports HTTPS (Exp\#8) and Web server Nginx (Exp\#9) with obvious better tail latency than F-Stack.

\end{itemize}
%-------------------------------------------------------------------------------
\subsection{Setup} 
\textbf{Testbed.}
We deployed QStack prototype in a dual-socket machine 
with Intel(R) Xeon(R) E5-2630 v4 $2.2$GHz, $64$\,GB\ RAM, 
and an Intel 82599ES $10$\,Gbps\ NIC. 
We only evaluate the first socket without hyper-threading, \textit{i.e.}. 8 cores, unless specified.

We used $2$ clients running workload generators and each uses a machine 
with $2\times$ Intel(R) Xeon(R) CPU E5645 2.4GHz, $32$\,GB\ RAM, and an Intel 82599ES $10$\,Gbps\ NIC. 
The workload generators and the server are connected with a HW-C16800 switch. 

\textbf{Workloads.} We study the load characteristics of front-end services in the data center and find that there are two kinds of classical modes: 1) massive short-running requests, which can be processed at the microsecond level, such as the distributed memory cache system (MICA), or the front-end server in the microservice architecture that simply analyzes requests and forwards them to other servers for processing (router service in the microservice benchmark usuit); 2) A large number of short-running requests, mixed with complex requests ~\cite{rhythm-eurosys-2020} with execution time gap to 2-3 orders of magnitude. For example, in the memory database RocksDB, a large number of get/set requests with execution time of microseconds are mixed with scan requests with processing time of milliseconds; Facebook USR dataset ~\cite{fbworkkoad-2012} is of similar features, 99\% requests is get, and the request processing is very short.
In combination with the above analysis of the data center load, we use MCC (Massive Client Connections) ~\cite{mcc-2019} to simulate the load of the above two modes: 1) simple requests, and the execution time of each request is 1 microsecond or 10 microseconds; 2) A large number of simple requests are mixed with a small number of complex requests, consisting of 99.5\% requests with 1 microsecond execution time and 0.5\% requests with 1 millisecond execution time. Here, we have amplified the proportion of requests with different priorities and the difference in execution time to highlight the interference effect.
Further we adopt MCC to generate complicated workload for application test. And to be more convincing, we tested Nginx with the popular Web pressure tool wrk~\cite{WEB:wrk}. 

\begin{figure}[t]
  \centering 
  \subfigure [Taking 60s as statistical period
  (idle: core without any running coroutines;
  A: core running application coroutine only; 
  S: core running network stack coroutine only; 
  A+S: core shared by application coroutine and network stack coroutine
  )] { 
    \includegraphics[width=0.66\linewidth]{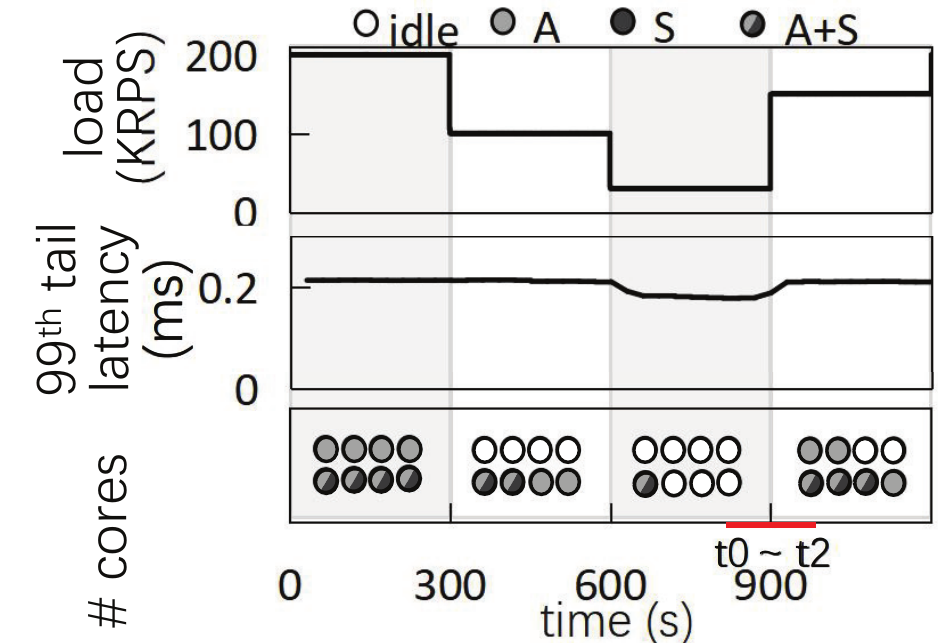}
    \label{fig:evaluation_elasticframework1}
  } 
  \centering 
  \subfigure [Taking 10ms as statistical period at load rising edge] { 
    \includegraphics[width=0.66\linewidth]{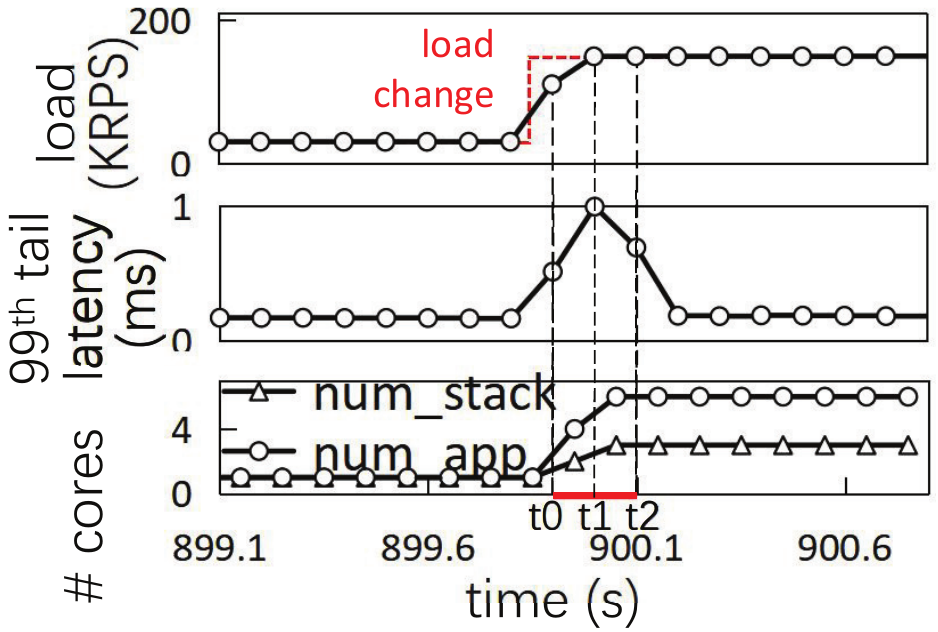}
    \label{fig:evaluation_elasticframework2}
  } 
  %\vspace{0.5cm}
  \caption{
    Resource dynamic adjustment 
  }
  %\vspace{0.6cm}    
  \label{fig:evaluation_elasticframework}
\end{figure}

\textbf{Methodology.} 
We compare QStack to Shenango, mTCP, IX and F-Stack. Shenango\ also operates in the first socket without hyper-threading, as we found that there is a significant performance drop with hyper-threading for Shenango. We use one IOKernel in Shenango, but as Shenango\ cannot run worker's runtime on the sibling core of IOKernel, there are actually at most 6 active cores being used for applications.
For all experiments involving Shenango, we have to keep the number of concurrent connections below 10,000, as Shenango\ is unable to complete all the connection establishments if the concurrency is over 12,000.
mTCP, IX\ and F-Stack\ runs on a fixed set of cores, 
\textit{i.e.}, 8 cores in our experiments. 
We evaluate the targeted system performance mainly with two metrics: 
(1) tail-latency, \textit{i.e.}, the $99^{th}$ percentile server-side latency 
measured the time between the query entering the server 
and its response leaving the server.
We use an open source monitoring tool to measure this~\cite{hcmonitor-CPE-21}.
and 
(2) CPU efficiency calculated as the proportion of CPU cycles used in application routines, denoted by $\eta$, 
\textit{i.e.}, $\eta = \gamma_{app} / \gamma_{total}$, 
where $\gamma_{total}$ denotes the cycles used by both application coroutines 
and network coroutines including overhead like spin-polling and context switch, 
$\gamma_{app}$ denotes the cycles used for typical service in application coroutines.

\paragraph{(Exp\#1)\textbf{Resource dynamic detect and adjustment.}}

In Figure~\ref{fig:evaluation_elasticframework}, we show that QStack can dynamically detect and adjust network stack coroutines and application coroutines according to real-time load. In particular, the load change moment around 900 milliseconds was measured in detail. For convenience, here we take 200KRPS as 100\% load reference. At time t0 for the monitor, the load increased, and since QStack has not made CPU adjustment (only one core is used), the response tail latency increased. Then the resource manager calculates the previous statistical period load, and obtains the average load from 10\% (20 KRPS) to 75\%. Adjust the number of network stack coroutines and application coroutines to 2 and 4. However, from t0 to t1, the real-time load has stabilized at 75\% (150 KRPS). The insufficient CPU resources allocated by QStack led to a significant increase in the tail latency detected by the monitor at t1. At t1-t2 detect point, the resource manager detected that the load in the last statistical period T reached 75\%, and adjusted the number of network stack coroutines and application coroutines to 3 and 6. At the tail latency statistical point t2 for monitor, the tail latency  starts to decrease, and recovers to the normal level in the next statistical period.
The experimental results show that after the load change, the network stack can timely detect the load change within two statistical periods (less than 2T, here it is 200ms from t0 to t2), schedule the CPU resources, and then the response tail latency goes back to the normal level (the response tail latency has decreased after t2, but the monitor needs one more statistical period to get).

\begin{figure}
  \begin{minipage}[!htbp]{0.48\linewidth}
    \centerline{\includegraphics[width=4cm]{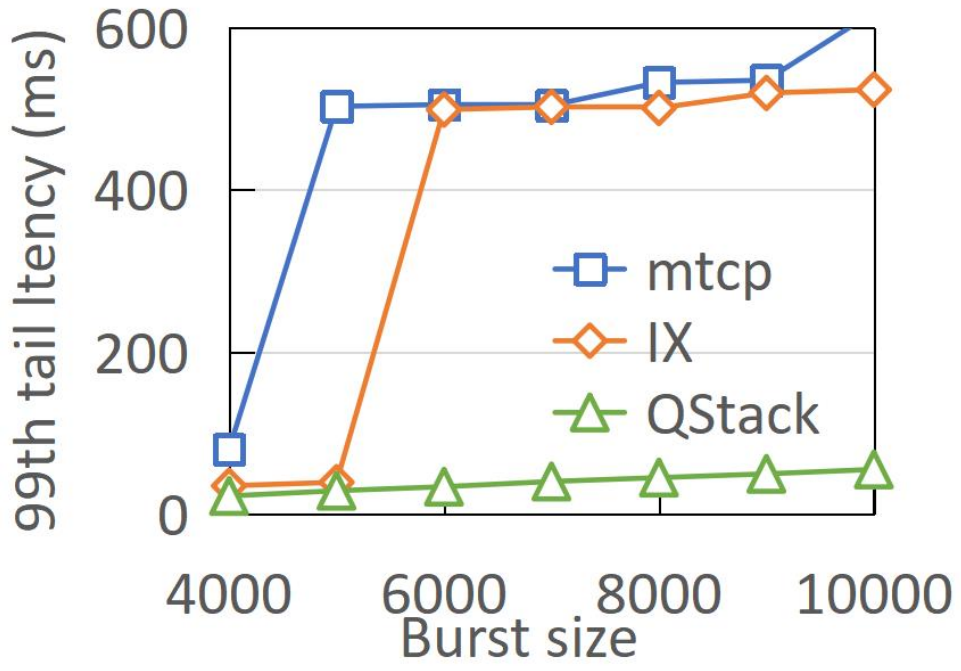}}
    %\vspace{-0.3cm}
    \caption{Fastcalldown to avoid packet drop in NIC.}
    \label{fig:evaluation_packetdrop}
  \end{minipage}
  \hspace{0.1cm}
  \begin{minipage}[!htbp]{0.48\linewidth}
    \centerline{\includegraphics[width=4cm]{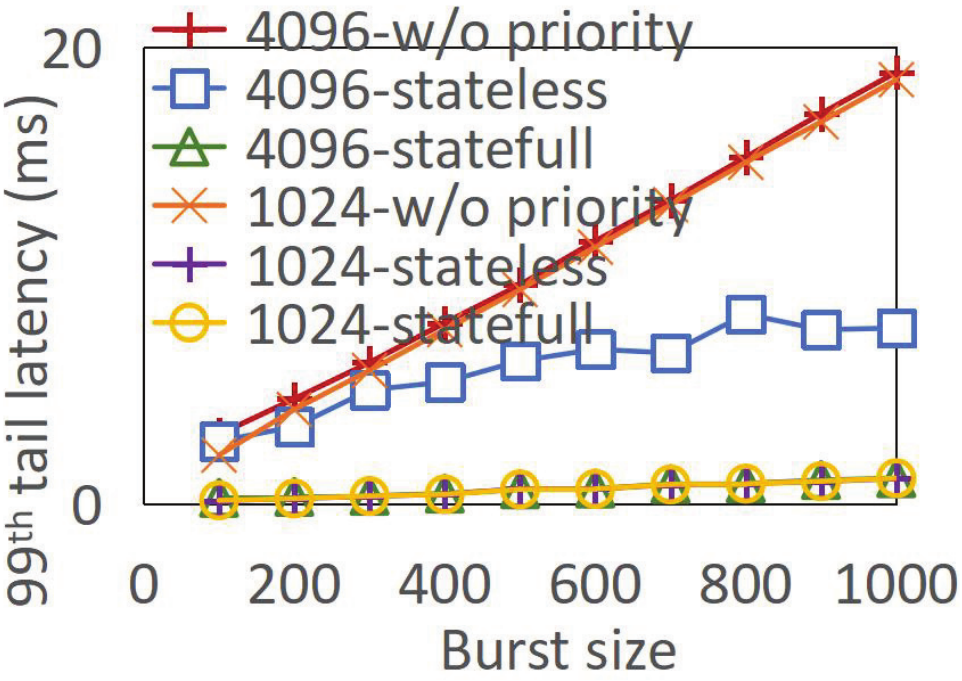}}
    %\vspace{-0.3cm}
    \caption{Application defined priority for cross-packet message in TCP layer}
    \label{fig:evaluation_App_information_analysis}
  \end{minipage}
  %\vspace{-0.6cm}
\end{figure}

\paragraph{(Exp\#2)\textbf{Fastcalldown to avoid packet drop in NIC.}}

With the support of fastcalldown and application-semantic analysis,
QStack provides both low latency and high CPU efficiency 
when co-locating multiple applications with different types on the same server. We use MCC to generate 99.5\% of 1 microsecond short-running requests and 0.5\% of 1 millisecond long-running requests, with a burst rate of 1 burst per second.

As shown in Figure~\ref{fig:evaluation_packetdrop}, the response tail latency of mTCP and IX suddenly increases to 500 milliseconds. This is because a large number of packets reach the network card when the application coroutine occupies the processor core for a long time. Packet drop is happened by NIC queue overflow, resulting in an delay increase caused by timeout retransmission. Benefited from the fastcalldown design, QStack can receive packets from the NIC queue in time to avoid packet drop, thus keeping the tail latency at a low level.

\paragraph{(Exp\#3)\textbf{Application defined priority for cross-packet message in TCP layer.}}

In this experiment, we verify the role of using TCP flow private field to analyze stateful scheduling feature across packets in application defined transport layer scheduling feature extraction. We use MCC to generate mixed priority requests. The execution time of each request is 100 microseconds. We design a random 5\% high priority delay sensitive request, and count the 99th response tail latency of this type request. This load can simulate that the data center server receives an external Web request. The request length is slightly longer due to the user data, and services of different service quality levels are provided according to the request type, user ID, etc. We conducted two experiments. The request length is 1024 bytes (1 TCP packet transmission required) and 4096 bytes (3 TCP packet transmission required). The request length is written in the request header. On the server side, we use the TCP private field to implement stateful application definition and transport layer scheduling feature extraction, and judge the scheduling priority according to the request type in the request header. When the request length exceeds the upper limit of a single TCP packet, the scheduling feature extraction callback function records the current number of unread data bytes and the corresponding scheduling type in the TCP private field, And mark the subsequent packets belonging to the TCP flow request as the same scheduling type. We compared the use of stateless applications to define the driver layer scheduling feature extraction. When the request length exceeds a TCP packet length, the driver layer scheduling feature extraction callback function can only identify the first packet scheduling type, and mark the subsequent packets of the request as the default scheduling priority.

As shown in Figure~\ref{fig:evaluation_App_information_analysis}, when the request length is 4096 bytes, because the stateless driver layer cannot correctly extract the scheduling feature of the next few packets, delay sensitive requests are affected by the queuing delay. Qstack, which uses TCP private field for stateful transport layer scheduling feature extraction, can still maintain effective scheduling feature extraction and scheduling for 4096 byte delay sensitive requests, and its response tail latency is 15\% of the stateless. It is also better than unlabeled scheduling. This is because when the upper layer application receives a high priority packet receiving event, it will try to receive all the data of the current TCP connection. If all the requested packets are already in the TCP receiving buffer, the high priority request can be processed immediately. The 1024 byte request experiment results are as references.

\paragraph{(Exp\#4)\textbf{Priority under hybrid deployed applications.}}

\begin{figure*}[t]
  \centering 
  \subfigure{
      \includegraphics[width=16cm]{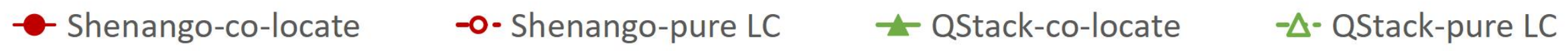}
  }
  %\vspace {-0.5cm}
  \setcounter{subfigure}{0}
  \subfigure [LC latency with 20 burst/sec] { 
    \includegraphics[width=4cm]{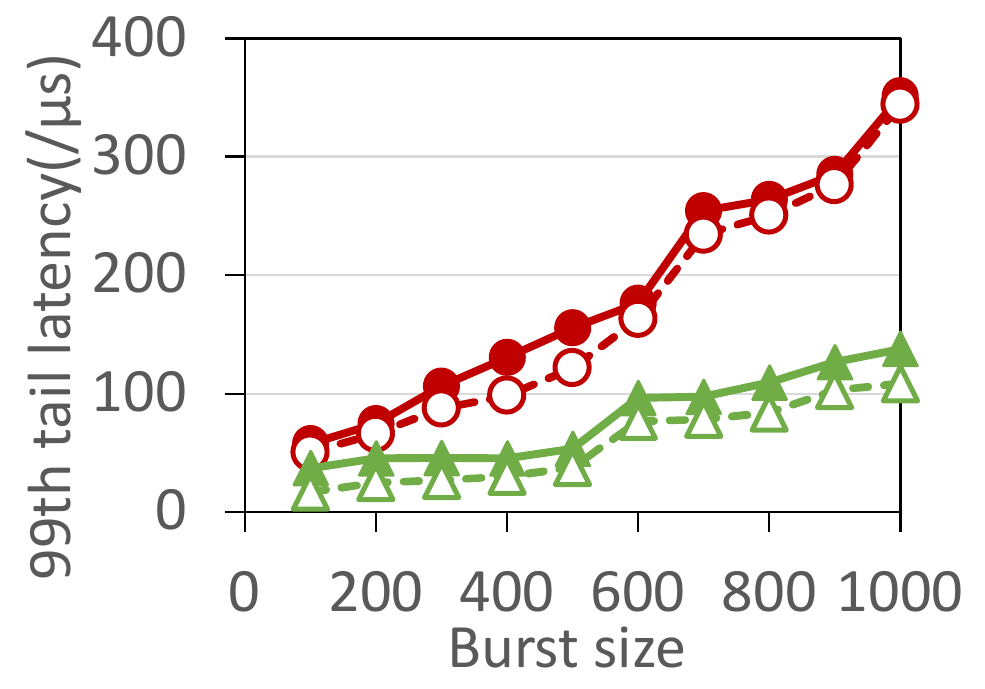}
    \label{fig:evaluation-mixed_latency_low}
  } 
  \subfigure [CPU efficiency with 20 burst/sec] { 
    \includegraphics[width=4cm]{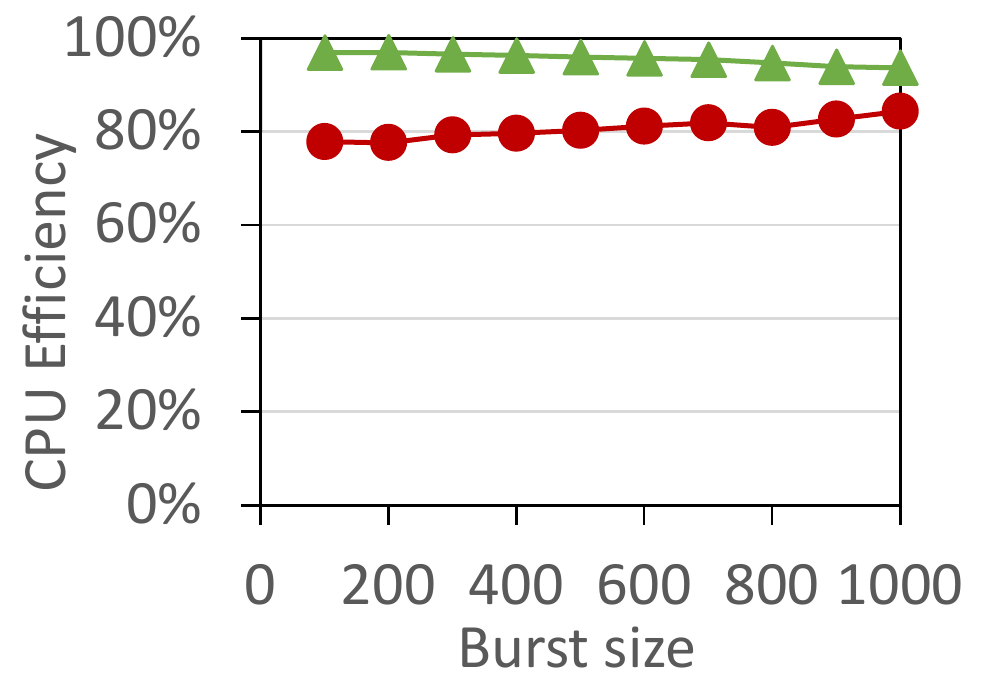}
    \label{fig:evaluation-mixed_throughput_low}
  } 
  \subfigure [LC latency with 100 burst/sec] { 
    \includegraphics[width=4cm]{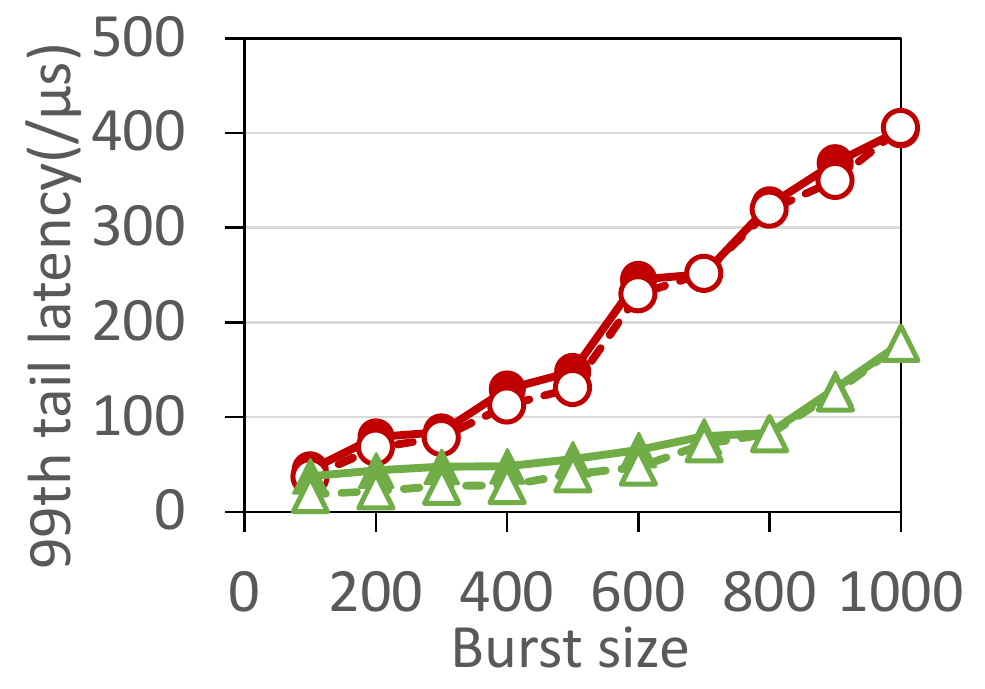}
    \label{fig:evaluation-mixed_latency_high}
  }
  \subfigure [CPU efficiency with 100 burst/sec] { 
    \includegraphics[width=4cm]{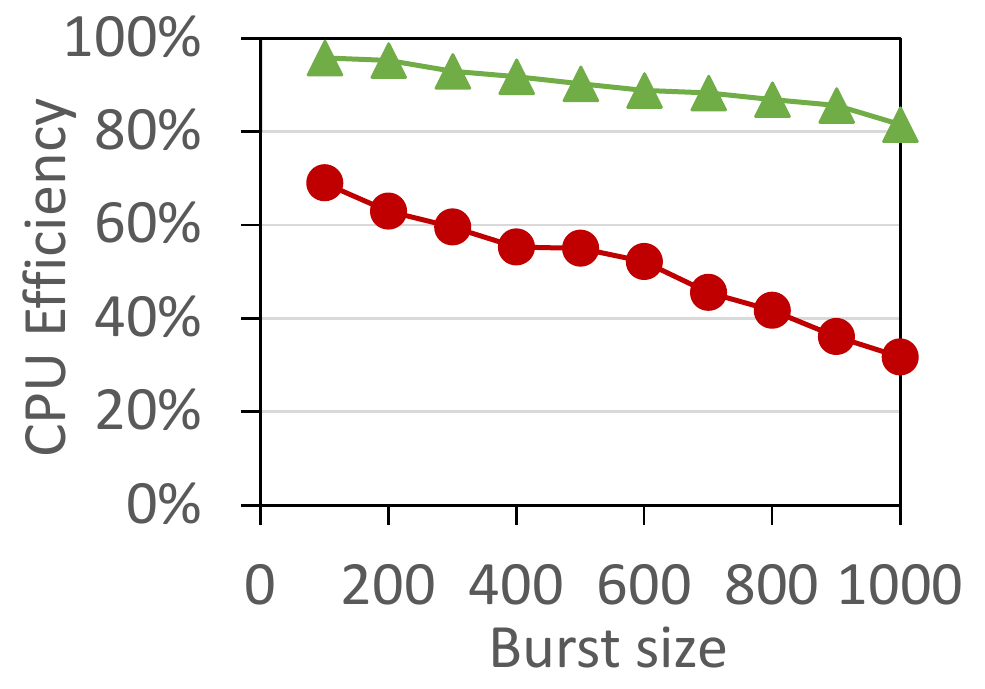}
    \label{fig:evaluation-mixed_throughput_high}
  }

  \caption{%\vspace
  tail latency and CPU efficiency when co-locating \textit{Latency-Critical} service with \textit{Best-Effort} service
  }
  (pure LC: only latency critical application 
  ; co-locate: both latency critical and best-effort application co-located) 
  \label{fig:evaluation-colocate}
\end{figure*}

With the support of fastcalldown and application-semantic analysis,
QStack can provide both low latency and high CPU efficiency 
when co-locating multiple applications with different types on the same server.

In this experiment, we co-locate Latency-Critical (LC) application and Best-Effort (BE) service both generated by MCC over QStack. We first measure the tail latency of the LC service without running BE service, 
and then measure the their co-locating, 
and evaluate the CPU efficiency with the matrix.
We use two burst rates: $200$ bursts/sec and $1000$ bursts/sec, 
and vary the burst size from 100 to 1000, to demonstrate the workload with different network load. 

As shown in Figure~\ref{fig:evaluation-colocate}, 
QStack provides both lower tail-latency and higher CPU efficiency than Shenango. 
Figure~\ref{fig:evaluation-mixed_latency_low} and Figure~\ref{fig:evaluation-mixed_latency_high}
show that, there is little interference on the tail-latency of LC service 
from BE service for both QStack and Shenango, 
while QStack outperforms Shenango\ on tail-latency by more than $2\times$ 
for its low overhead. 
Figure~\ref{fig:evaluation-mixed_throughput_low} and Figure~\ref{fig:evaluation-mixed_throughput_high} show that 
QStack always achieves better CPU efficiency than Shenango\ 
by utilizing all cores, while Shenango dedicated cores for dispatching, which cannot be used by applications.
 
Moreover, when the load of LC service increases in 
Figure~\ref{fig:evaluation-mixed_throughput_high}, the CPU efficiency of Shenango drops significantly, which is $50$\%\ lower than QStack, due to more context switch.

\paragraph{(Exp\#5)\textbf{Application defined priority in driver layer\\}}

In this experiment, we test the impact of application defined driver scheduling feature extraction on the tail latency of delay sensitive request. We use MCC to generate mixed priority requests. The processing time of each request is 10 microseconds. We design random 5\% requests as high priority and 95\% as low priority. We statistic the 99th response tail latency of high priority requests. This load can be used to simulate RPC services in the data center, and distinguish the request service level according to the request type, user ID, etc. On the server side, we use the application definition driver layer scheduling feature extraction to identify the packet scheduling feature, and compare it with the application-defined transport layer scheduling feature extraction.

\begin{figure}
  \begin{minipage}[!htbp]{0.48\linewidth}
    \centerline{\includegraphics[width=4cm]{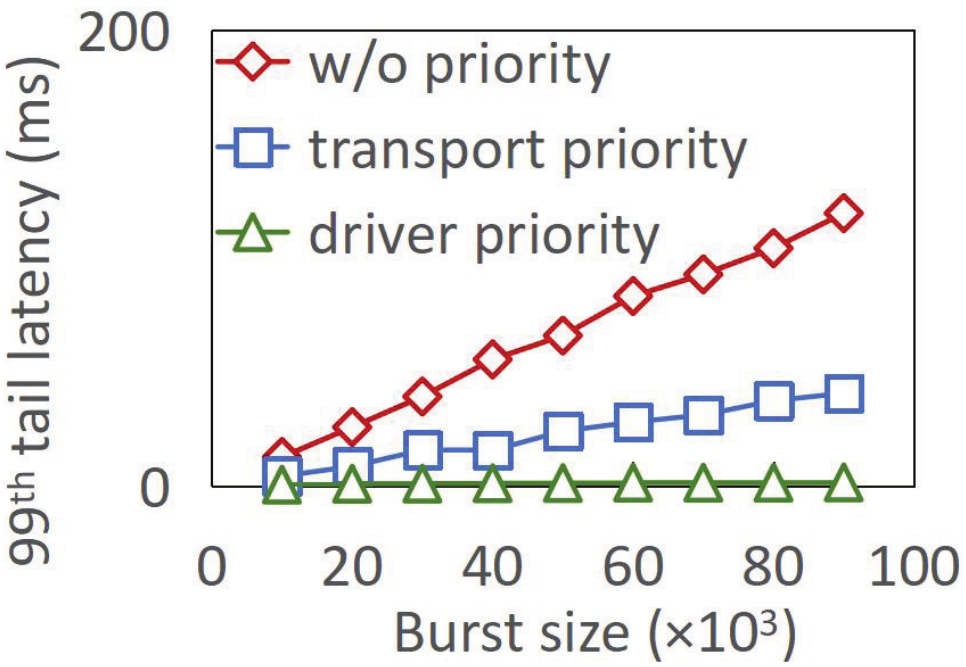}}
    %\vspace{-0.3cm}
    \caption{Application defined priority in driver layer}
    \label{fig:evaluation_App_informatio_extraction}
  \end{minipage}
  \hspace{0.1cm}
  \begin{minipage}[!htbp]{0.40\linewidth}
    \centerline{\includegraphics[width=4cm]{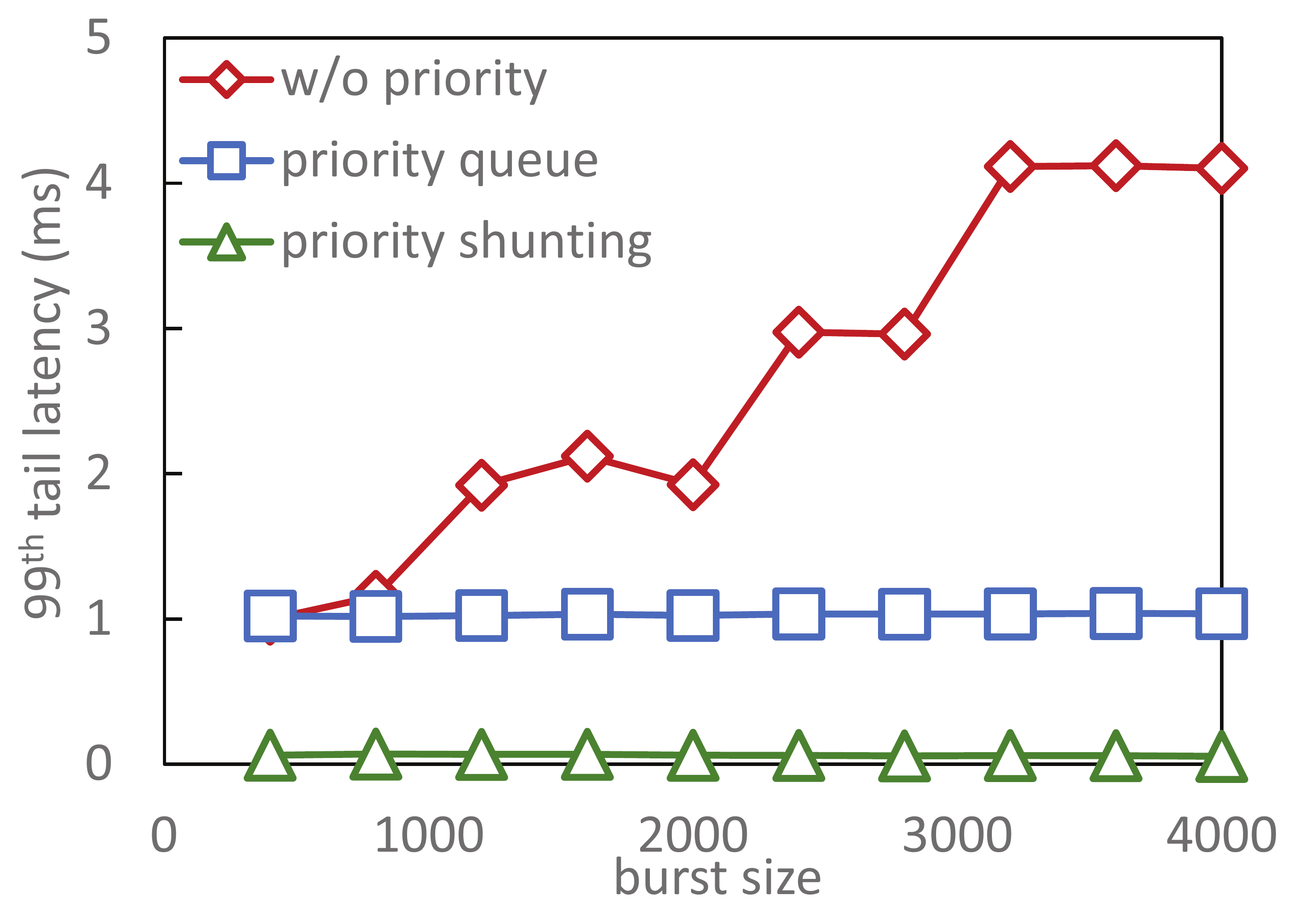}}
    %\vspace{-0.3cm}
    \caption{Tail latency of priority diffluence based on elastic framework.}
    \label{fig:evaluation_pri_km}
  \end{minipage}
  %\vspace{-0.6cm}
\end{figure}

As shown in Figure~\ref{fig:evaluation_App_informatio_extraction}, when using the application defined driver layer to extract scheduling feature, since the queuing delay of high priority requests is effectively avoided, the response tail latency of high priority requests is reduced by 98.5\% compared with that without labeled scheduling. Although the application defined transport layer scheduling feature extraction implements priority queue scheduling in the event framework to avoid the queuing delay in the application layer, high priority requests will still experience queuing delay when the packet receiving buffer waits for the network stack to process. Therefore, the response tail latency of high priority requests is 24.5 times higher than that when the application defined driver layer scheduling feature extraction is used.
\\

\paragraph{(Exp\#6)\textbf{Priority diffluence.}}
Using MCC to generate mixed type requests. The processing time of 99.5\% requests is 1 microsecond, and the processing time of 0.5\% requests is 1 millisecond. Measure and count up the response tail latency of 1 microsecond requests. Priority queue scheduling avoids the blocking delay in most cases and greatly reduces the response delay. However, since the application coroutine starts to process low priority requests for 1 millisecond after processing high priority requests, subsequent high priority requests would be blocked, so the response tail latency remains at 1 millisecond.
However, as shown in Figure~\ref{fig:evaluation_pri_km}, when collaboration with the elastic framework diffluence based on request feature labels and K:M coomunication, the low priority 1ms requests are able to be processed on a separate processor core, which will not have any interfere on the processing of high priority requests at all, effectively reducing the response tail latency of high priority requests to 1.6\% than the case with no labeled scheduling.

\paragraph{(Exp\#7)\textbf{High concurrent IoT service.}}

\begin{figure}
  \begin{minipage}[t]{0.48\linewidth}
    \centerline{\includegraphics[width=1 \linewidth]{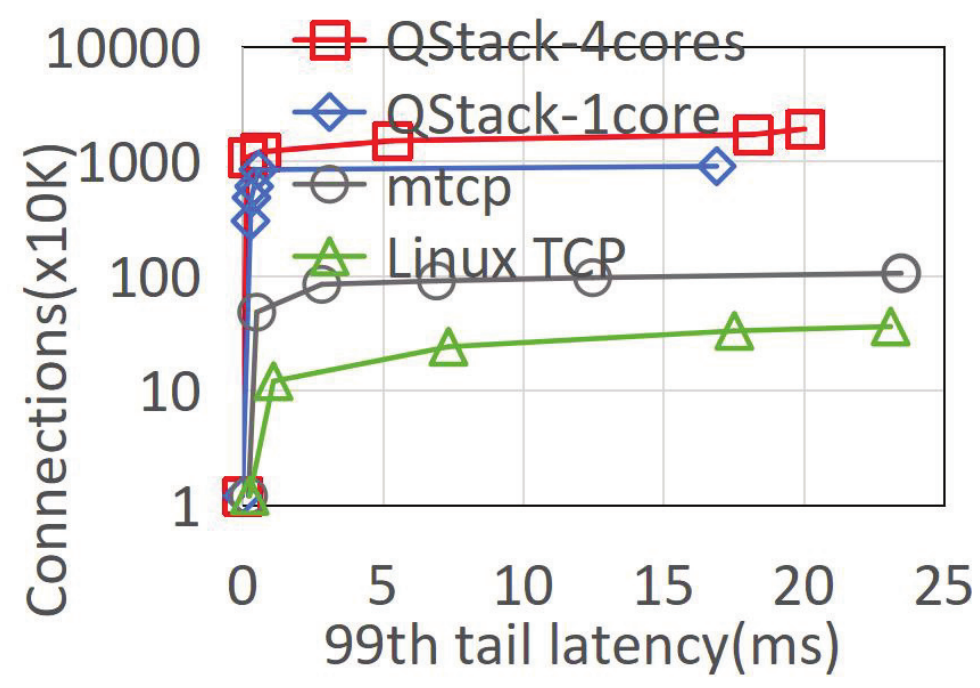}}
    %\vspace{-0.3cm}
    \caption{Test on high concurrent IoT service}
    \label{fig:evaluation-iotepserver}
  \end{minipage}
  \hspace{0.1cm}
  \begin{minipage}[t]{0.48\linewidth}
    \centerline{\includegraphics[width=1 \linewidth]{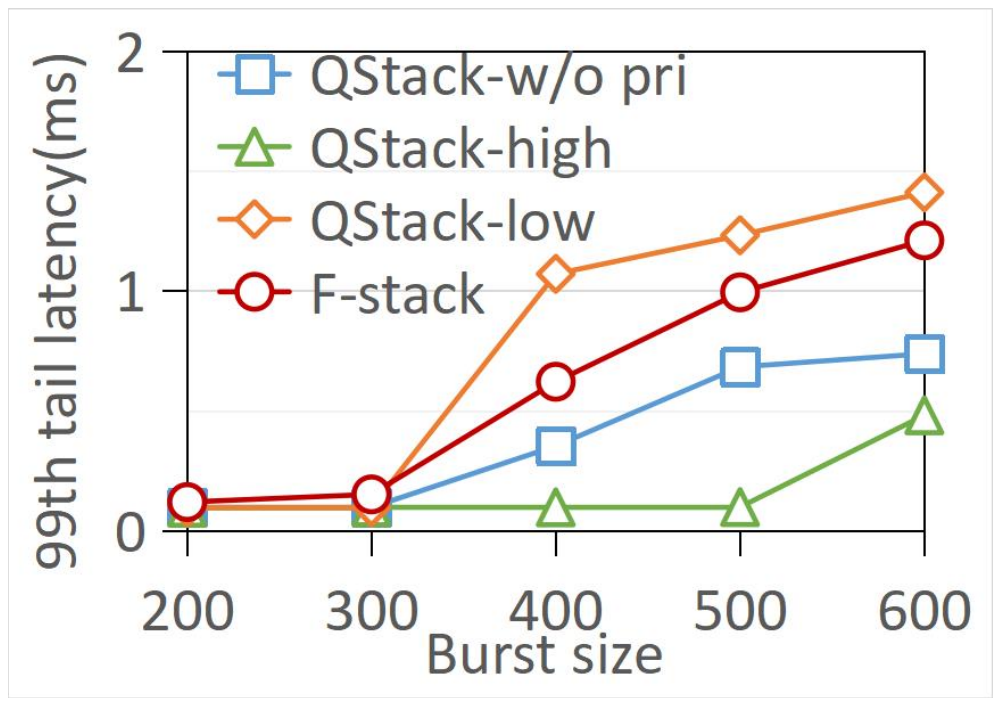}}
    %\vspace{-0.3cm}
    \caption{Test on priority supported HTTPs}
    \label{fig:evaluation-ssl_echo}
  \end{minipage}%
  %\vspace{-0.6cm}
\end{figure}

In this experiment, in order to test the performance of each network stack in handling high concurrent loads, we used MCC to simulate the Internet of Things traffic with high concurrency and high burst, compressed the traffic per second to 100ms to form bursts, and tested the trend on 99th response latency and concurrency. As can be seen from the Figure~\ref{fig:evaluation-iotepserver}, when 25ms is taken as the tail latency threshold, the maximum concurrent connection that Linux TCP can reach is only 0.36 million, mTCP can be increased to 1.05 million, while Qstack with one-core network stack can efficiently handle 9 million concurrent connections with the tail latency about 16 ms. While Qstack with four-core network stack can support 19 million concurrency, and the tail latency is only 20.03 ms. Therefore, compared with traditional Linux TCP and user-space network mTCP, Qstack has greatly improved its ability to handle high concurrency. The concurrency performance of Qstack with one-core network stack is more than 25 times that of Linux TCP and more than 8.5 times that of mTCP, and Qstack with four-core network stack can support more than 10 million concurrency in a single server.\\ 

\paragraph{(Exp\#8)\textbf{Priority supported HTTPs.}}

We ported HTTP\_TLS to QStack and F-Stack. To evaluate them, we used workload generated by MCC, where HTTP GET is usually delay critical, classified as high-priority by QStack, and POST does not need to response quickly, classified as low-priority. We vary the the burst size from $200$ to $600$ with burst rate 10/sec. The average service time of HTTPS queries is about $20$\mbox{$\mu s$}. We compare QStack with F-Stack~\cite{WEB:F-stack}, 
an open-source user-space network stack developed and used in Tencent.
As shown in Figure~\ref{fig:evaluation-ssl_echo}, 
we plotted tail-latency performances achieved by 
both high-priority queries (denoted by QStack-high) 
and low-priority queries (denoted by QStack-low). 
QStack with prioritized scheduling achieves the best tail-latency for high-priority queries, which is $0.6\times$ lower than F-Stack. 
QStack with event prioritization reduces $35$\%\ tail-latency of high-priority queries, compared to QStack without events prioritization 
(denoted by QStack-w/o-prio). 
The cost of prioritized scheduling is that the tail-latency of low-priority SSL queries is $1\times$ higher than the one achieved 
by QStack without event prioritization. 

\paragraph{(Exp\#9)\textbf{Web server Nginx.}}

\begin{table}[h]
  \caption{HTTP queries served by Nginx}
  \vspace{0.0cm}
  \begin{center}
      \small
      \begin{tabular}{lcc}
      \hline
            & Throughput (MQPS) & 99th tail latency (ms) \\
      \hline
  QStack  & 5.73              & 0.95              \\
  F-stack   & 5.29              & 3.07              \\
      \hline
      \end{tabular}
  \end{center}
  \label{tab:evaluation-nginx}
  \vspace{0.0cm}
  \end{table}

We evaluated the Nginx\ Web service on QStack\ 
using stress test mode HTTP GET workload generated by wrk~\cite{WEB:wrk}.
We compare Nginx\ performance on QStack\ with Nginx\ on F-Stack, 
and use $360$ long-lived TCP connections, 
which achieves the maximum throughput for both QStack and F-Stack. 
As shown in Table~\ref{tab:evaluation-nginx}, 
QStack achieves $1.1\times$ throughput of Nginx\ compared to F-Stack, 
while it reduces $67$\%\ tail-latency at the same time.

%-------------------------------------------------------------------------------
%\section*{related-work}
%-------------------------------------------------------------------------------
% \input{subsections/related-work}
\section{Related Work}
\label{sec:bg-relatedwork}

\textbf{CPU efficiency.}
Many systems bypass the kernel to achieve low-latency networking with user-space packet I/O libraries like DPDK or netmap~\cite{netmap-atc-2012}, and rebuild network stack in user space. Examples include lwIP~\cite{lwip-2001}, mTCP~\cite{mtcp-2014}, IX~\cite{ix-2014}, Arrakis~\cite{arrakis-2014}, MICA~\cite{mica-2014}, Sandstorm~\cite{sandstorm}, Zygos~\cite{zygos-2017}, Shinjuku~\cite{shinjuku-2019}, Shenango~\cite{shenango-2019}, Caladan~\cite{calcadan-osdi-2020}, LOS~\cite{los-apnet-2017} F-stack~\cite{WEB:F-stack}, SeaStar~\cite{WEB:SeaStar},  SocksDirect~\cite{socksdirect-2019}, eRPC~\cite{erpc-2019}, Snap~\cite{snap-sosp-2019}. 
User-space network stacks allocate CPUs in different ways. 
Shinjuku and Perséphone\cite{shinjuku-2019,idle-sosp-2021} use a fixed number of cores for network processing, \textit{e.g.}, 1 core in their implementation, and scheduling the application processing on the reset of cores with a centralized scheduler. The fixed core allocation will result in waste CPU cycles on polling at low load, and introduce a bottleneck at burst load. 
mTCP, IX, ZygOS, and F-Stack ~\cite{mtcp-2014,ix-2014,zygos-2017,WEB:F-stack} 
allocate cores ``symmetrically'' which pack one network thread and one application thread on each core, 
fixing the processing of each flow within a certain core at both network and application layer. 
This design provides good multi-core scalability and data locality. 
However, they cannot allocate resources on demand based on their 
respective loads, 
and could easily introduce NIC packet drop when the application occupies the CPU cycles for a long period. 
Shenango~\cite{shenango-2019} decouples NIC polling from network stack, 
uses one dedicated core for NIC polling, and dispatches packets to the other cores performing protocol and application processing. 
However, as the CPU used for NIC polling is fixed, again, 
the dedicated core for NIC polling will become the bottleneck at peak load, 
and will waste CPU cycles at low load. 
\textbf{The fundamental problem is that they do not dynamically adjust the CPU resource for network stack.}

\textbf{Service quarlity.}
Many works try to improve server efficiency without impact on response latency. Examples include PerfIso~\cite{perfiso-atc18}, Shenango~\cite{shenango-2019}, Rhythm~\cite{rhythm-eurosys-2020}, Arachne~\cite{arachne-osdi-2018}. 
In mTCP and IX~\cite{mtcp-2014,ix-2014}, there is no scheduling because 
every request is mapped to a core permanently for processing. 
This can introduce HoL blocking 
when short-running requests are waiting behind long-running requests 
that occupy the core, and thus experience a long queuing delay.
ZygOS and Shenango~\cite{zygos-2017} first dispatch requests to work queues of cores blindly, and the idle cores can steal requests from busy cores which might have been blocked by long-running requests, \textit{i.e.}, work-stealing.
Although it alleviates the HoL blocking to some extent, it introduces the extra overhead caused by locks and competition among cores. 
Moreover, the HoL blocking still exists when all cores are occupied by 
long-running requests. Shinjuku~\cite{shinjuku-2019} also firstly dispatches requests blindly, and uses a centralized user-level thread scheduler to preempt 
the application threads which have been running for a long period, 
\textit{e.g.}, more than $5$$\mbox{$\mu s$}$. However, preemption introduces high overhead, \textit{e.g.}, 2$\mbox{$\mu s$}$ overhead at a scheduling frequency of 5$\mbox{$\mu s$}$\cite{idle-sosp-2021}. The above schedulings are all performed in a reactive way, \textit{i.e.}, first dispatch the network requests indifferently, 
then adjust the CPU resource later according to application status. 
\textbf{The fundamental problem is that they ignore the application intrinsic characteristics, \textit{e.g.} short-running or long-running, priority, SLO, \textit{etc.}} 
This suggests we find a way to extract those characteristics in advance (\S~\ref{sec:bg-schedule}).

%-------------------------------------------------------------------------------
%\section*{Conclusion}
%-------------------------------------------------------------------------------
% \input{subsections/conclusion.tex}
%-------------------------------------------------------------------------------

\section{Conclusion}
QStack is a user-space TCP/IP network stack with elasticity and priority for diversified service demands (i.e., high concurrency, high CPU efficiency, low tail latency) in data center front-end server. Its design principle is to re-architect the network stack collaboration among CPU cores horizontally and the coordination across network layers vertically. We show how QStack achieves both CPU resource efficiency and service quality, with packet drop avoidance guarantee. Experiments show the effectiveness of QStack over 4 state-of-the-art user-space network stack designs.

\clearpage
%-------------------------------------------------------------------------------
% \bibliographystyle{plain}
\bibliographystyle{unsrt}
\bibliography{refer}

%%%%%%%%%%%%%%%%%%%%%%%%%%%%%%%%%%%%%%%%%%%%%%%%%%%%%%%%%%%%%%%%%%%%%%%%%%%%%%%%
\end{document}

%% file: defs.tex
\definecolor{pink}{rgb}{1.0,0.47,0.6}

\def\summary#1{}
\def\chsumm#1{}

\def\hide#1{}
\def\eat#1{}
\newcounter{mycomment}

%% file: paper.bbl
\begin{thebibliography}{10}

\bibitem{iot-2016}
Alessio Botta, Walter de~Donato, Valerio Persico, and Antonio Pescap{\`{e}}.
\newblock Integration of cloud computing and internet of things: {A} survey.
\newblock {\em Future Gener. Comput. Syst.}, 56:684--700, 2016.
\newblock \url{https://doi.org/10.1016/j.future.2015.09.021}.

\bibitem{WEB:idc-iot}
Idc.(2016) internet of things market statistics. [online].
\newblock
  \url{https://www.ironpaper.com/webintel/articles/marketing-opportunities-for-the-internet-of-things}.

\bibitem{frontend-imc-11}
Yingying Chen, Sourabh Jain, Vijay~Kumar Adhikari, and Zhi{-}Li Zhang.
\newblock Characterizing roles of front-end servers in end-to-end performance
  of dynamic content distribution.
\newblock In Patrick Thiran and Walter Willinger, editors, {\em Proceedings of
  the 11th {ACM} {SIGCOMM} Internet Measurement Conference, {IMC} '11, Berlin,
  Germany, November 2-, 2011}, pages 559--568. {ACM}, 2011.
\newblock \url{https://doi.org/10.1145/2068816.2068868}.

\bibitem{arrakis-2014}
Simon Peter, Jialin Li, Irene Zhang, Dan R.~K. Ports, Doug Woos, Arvind
  Krishnamurthy, Thomas~E. Anderson, and Timothy Roscoe.
\newblock Arrakis: The operating system is the control plane.
\newblock In {\em 11th {USENIX} Symposium on Operating Systems Design and
  Implementation, {OSDI} '14, Broomfield, CO, USA, October 6-8, 2014.}, pages
  1--16, 2014.

\bibitem{mtcp-2014}
Eunyoung Jeong, Shinae Woo, Muhammad~Asim Jamshed, Haewon Jeong, Sunghwan Ihm,
  Dongsu Han, and KyoungSoo Park.
\newblock mtcp: a highly scalable user-level {TCP} stack for multicore systems.
\newblock In {\em Proceedings of the 11th {USENIX} Symposium on Networked
  Systems Design and Implementation, {NSDI} 2014, Seattle, WA, USA, April 2-4,
  2014}, pages 489--502, 2014.
\newblock
  \url{https://www.usenix.org/conference/nsdi14/technical-sessions/presentation/jeong}.

\bibitem{2013c10m}
Robert Graham.
\newblock The secret to 10 million concurrent connections -the kernel is the
  problem, not the solution.
\newblock 2013.
\newblock
  \url{http://highscalability.com/blog/2013/5/13/the-secret-to-10-million-concurrent-connections-the-kernel-i.html}.

\bibitem{shinjuku-2019}
Kostis Kaffes, Timothy Chong, Jack~Tigar Humphries, Adam Belay, David
  Mazi{\`{e}}res, and Christos Kozyrakis.
\newblock Shinjuku: Preemptive scheduling for {\(\mu\)}second-scale tail
  latency.
\newblock In {\em 16th {USENIX} Symposium on Networked Systems Design and
  Implementation, {NSDI} 2019, Boston, MA, February 26-28, 2019.}, pages
  345--360, 2019.
\newblock \url{https://www.usenix.org/conference/nsdi19/presentation/kaffes}.

\bibitem{idle-sosp-2021}
Henri~Maxime Demoulin, Joshua Fried, Isaac Pedisich, Marios Kogias, Boon~Thau
  Loo, Linh Thi~Xuan Phan, and Irene Zhang.
\newblock When idling is ideal: Optimizing tail-latency for heavy-tailed
  datacenter workloads with pers{\'{e}}phone.
\newblock In {\em {SOSP} '21: {ACM} {SIGOPS} 28th Symposium on Operating
  Systems Principles, Virtual Event / Koblenz, Germany, October 26-29, 2021},
  pages 621--637, 2021.
\newblock \url{https://doi.org/10.1145/3477132.3483571}.

\bibitem{shenango-2019}
Amy Ousterhout, Joshua Fried, Jonathan Behrens, Adam Belay, and Hari
  Balakrishnan.
\newblock Shenango: Achieving high {CPU} efficiency for latency-sensitive
  datacenter workloads.
\newblock In {\em 16th {USENIX} Symposium on Networked Systems Design and
  Implementation, {NSDI} 2019, Boston, MA, February 26-28, 2019}, pages
  361--378, 2019.
\newblock
  \url{https://www.usenix.org/conference/nsdi19/presentation/ousterhout}.

\bibitem{ix-2014}
Adam Belay, George Prekas, Ana Klimovic, Samuel Grossman, Christos Kozyrakis,
  and Edouard Bugnion.
\newblock {IX:} {A} protected dataplane operating system for high throughput
  and low latency.
\newblock In {\em 11th {USENIX} Symposium on Operating Systems Design and
  Implementation, {OSDI} '14, Broomfield, CO, USA, October 6-8, 2014.}, pages
  49--65, 2014.
\newblock
  \url{https://www.usenix.org/conference/osdi14/technical-sessions/presentation/belay}.

\bibitem{zygos-2017}
George Prekas, Marios Kogias, and Edouard Bugnion.
\newblock Zygos: Achieving low tail latency for microsecond-scale networked
  tasks.
\newblock In {\em Proceedings of the 26th Symposium on Operating Systems
  Principles, Shanghai, China, October 28-31, 2017}, pages 325--341, 2017.
\newblock \url{https://doi.org/10.1145/3132747.3132780}.

\bibitem{WEB:F-stack}
F-stack.
\newblock \url{http://www.f-stack.org/}.

\bibitem{WEB:dpdk}
Intel dpdk.
\newblock \url{https://www.dpdk.org/}.

\bibitem{microservice-asplos-2019}
Yu~Gan, Yanqi Zhang, Dailun Cheng, Ankitha Shetty, Priyal Rathi, Nayan Katarki,
  Ariana Bruno, Justin Hu, Brian Ritchken, Brendon Jackson, Kelvin Hu, Meghna
  Pancholi, Yuan He, Brett Clancy, Chris Colen, Fukang Wen, Catherine Leung,
  Siyuan Wang, Leon Zaruvinsky, Mateo Espinosa, Rick Lin, Zhongling Liu, Jake
  Padilla, and Christina Delimitrou.
\newblock An open-source benchmark suite for microservices and their
  hardware-software implications for cloud {\&} edge systems.
\newblock In {\em Proceedings of the Twenty-Fourth International Conference on
  Architectural Support for Programming Languages and Operating Systems,
  {ASPLOS} 2019, Providence, RI, USA, April 13-17, 2019}, pages 3--18, 2019.
\newblock \url{https://doi.org/10.1145/3297858.3304013}.

\bibitem{usuite-2018}
Akshitha Sriraman and Thomas~F. Wenisch.
\newblock {\(\mu\)} suite: {A} benchmark suite for microservices.
\newblock In {\em 2018 {IEEE} International Symposium on Workload
  Characterization, {IISWC} 2018, Raleigh, NC, USA, September 30 - October 2,
  2018}, pages 1--12, 2018.
\newblock \url{https://doi.org/10.1109/IISWC.2018.8573515}.

\bibitem{DBLP:conf/sigcomm/RoyZBPS15}
Arjun Roy, Hongyi Zeng, Jasmeet Bagga, George Porter, and Alex~C. Snoeren.
\newblock Inside the social network's (datacenter) network.
\newblock In {\em Proceedings of the 2015 {ACM} Conference on Special Interest
  Group on Data Communication, {SIGCOMM} 2015, London, United Kingdom, August
  17-21, 2015}, pages 123--137, 2015.
\newblock \url{https://doi.org/10.1145/2785956.2787472}.

\bibitem{traffic-in-the-wild-IMC-2010}
Theophilus Benson, Aditya Akella, and David~A. Maltz.
\newblock Network traffic characteristics of data centers in the wild.
\newblock In {\em Proceedings of the 10th {ACM} {SIGCOMM} Internet Measurement
  Conference, {IMC} 2010, Melbourne, Australia - November 1-3, 2010}, pages
  267--280, 2010.

\bibitem{microservice-overload-wechat-socc-2018}
Hao Zhou, Ming Chen, Qian Lin, Yong Wang, Xiaobin She, Sifan Liu, Rui Gu,
  Beng~Chin Ooi, and Junfeng Yang.
\newblock Overload control for scaling wechat microservices.
\newblock In {\em Proceedings of the {ACM} Symposium on Cloud Computing, SoCC
  2018, Carlsbad, CA, USA, October 11-13, 2018}, pages 149--161, 2018.
\newblock \url{https://doi.org/10.1145/3267809.3267823}.

\bibitem{tensorflow-osdi-2016}
Mart{\'{\i}}n Abadi, Paul Barham, Jianmin Chen, Zhifeng Chen, Andy Davis,
  Jeffrey Dean, Matthieu Devin, Sanjay Ghemawat, Geoffrey Irving, Michael
  Isard, Manjunath Kudlur, Josh Levenberg, Rajat Monga, Sherry Moore,
  Derek~Gordon Murray, Benoit Steiner, Paul~A. Tucker, Vijay Vasudevan, Pete
  Warden, Martin Wicke, Yuan Yu, and Xiaoqiang Zheng.
\newblock Tensorflow: {A} system for large-scale machine learning.
\newblock In {\em 12th {USENIX} Symposium on Operating Systems Design and
  Implementation, {OSDI} 2016, Savannah, GA, USA, November 2-4, 2016}, pages
  265--283, 2016.
\newblock
  \url{https://www.usenix.org/conference/osdi16/technical-sessions/presentation/abadi}.

\bibitem{DBLP:conf/sigcomm/Zhang0KGJ19}
Xu~Zhang, Siddhartha Sen, Daniar Kurniawan, Haryadi Gunawi, and Junchen Jiang.
\newblock {E2E:} embracing user heterogeneity to improve quality of experience
  on the web.
\newblock In {\em Proceedings of the {ACM} Special Interest Group on Data
  Communication, {SIGCOMM} 2019, Beijing, China, August 19-23, 2019}, pages
  289--302, 2019.
\newblock \url{https://doi.org/10.1145/3341302.3342089}.

\bibitem{lns-jcst-2020}
Wenli Zhang, Ke~Liu, Yifan Shen, Yazhu Lan, Hui Song, Mingyu Chen, and
  Yuan{-}Fei Chen.
\newblock Labeled network stack: {A} high-concurrency and low-tail latency
  cloud server framework for massive iot devices.
\newblock {\em J. Comput. Sci. Technol.}, 35(1):179--193, 2020.
\newblock \url{https://doi.org/10.1007/s11390-020-9651-x}.

\bibitem{2021The}
I.~Zhang, A.~Raybuck, P.~Patel, K.~Olynyk, J.~Nelson, O.~S. Leija, A.~Martinez,
  J.~Liu, A.~K. Simpson, and S.~Jayakar.
\newblock The demikernel datapath os architecture for microsecond-scale
  datacenter systems.
\newblock pages 195--211, 2021.

\bibitem{rhythm-eurosys-2020}
Laiping Zhao, Yanan Yang, Kaixuan Zhang, Xiaobo Zhou, Tie Qiu, Keqiu Li, and
  Yungang Bao.
\newblock Rhythm: component-distinguishable workload deployment in datacenters.
\newblock In Angelos Bilas, Kostas Magoutis, Evangelos~P. Markatos, Dejan
  Kostic, and Margo~I. Seltzer, editors, {\em EuroSys '20: Fifteenth EuroSys
  Conference 2020, Heraklion, Greece, April 27-30, 2020}, pages 19:1--19:17.
  {ACM}, 2020.
\newblock \url{https://doi.org/10.1145/3342195.3387534}.

\bibitem{fbworkkoad-2012}
Berk Atikoglu, Yuehai Xu, Eitan Frachtenberg, Song Jiang, and Mike Paleczny.
\newblock Workload analysis of a large-scale key-value store.
\newblock In {\em {ACM} {SIGMETRICS/PERFORMANCE} Joint International Conference
  on Measurement and Modeling of Computer Systems, {SIGMETRICS} '12, London,
  United Kingdom, June 11-15, 2012}, pages 53--64, 2012.
\newblock \url{https://doi.org/10.1145/2254756.2254766}.

\bibitem{mcc-2019}
Wenqing Wu, Xiao Feng, Wenli Zhang, and Mingyu Chen.
\newblock {MCC:} {A} predictable and scalable massive client load generator.
\newblock In {\em Benchmarking, Measuring, and Optimizing - Second BenchCouncil
  International Symposium, Bench 2019, Denver, CO, USA, November 14-16, 2019,
  Revised Selected Papers}, pages 319--331, 2019.
\newblock \url{https://doi.org/10.1007/978-3-030-49556-5\_29}.

\bibitem{WEB:wrk}
Wrk.
\newblock \url{https://github.com/wg/wrk}.

\bibitem{hcmonitor-CPE-21}
Song H, Zhang W, Liu Ke, Shen Y, and Chen M.
\newblock Hcmonitor: An accurate measurement system for high concurrent network
  services.
\newblock In {\em Concurrency and Computation practice and Experience.
  2021(2)}, 2021.

\bibitem{netmap-atc-2012}
Luigi Rizzo.
\newblock netmap: {A} novel framework for fast packet {I/O}.
\newblock In {\em 2012 {USENIX} Annual Technical Conference, Boston, MA, USA,
  June 13-15, 2012}, pages 101--112, 2012.
\newblock
  \url{https://www.usenix.org/conference/usenixsecurity12/technical-sessions/presentation/rizzo}.

\bibitem{lwip-2001}
Adam Dunkels.
\newblock Design and implementation of the lwip tcp/ip stack.
\newblock {\em Swedish Institute of Computer Science}, 2(77), 2001.

\bibitem{mica-2014}
Hyeontaek Lim, Dongsu Han, David~G. Andersen, and Michael Kaminsky.
\newblock {MICA:} {A} holistic approach to fast in-memory key-value storage.
\newblock In {\em Proceedings of the 11th {USENIX} Symposium on Networked
  Systems Design and Implementation, {NSDI} 2014, Seattle, WA, USA, April 2-4,
  2014}, pages 429--444, 2014.
\newblock
  \url{https://www.usenix.org/conference/nsdi14/technical-sessions/presentation/lim}.

\bibitem{sandstorm}
Ilias Marinos, Robert N.~M. Watson, and Mark Handley.
\newblock Network stack specialization for performance.
\newblock In {\em {ACM} {SIGCOMM} 2014 Conference, SIGCOMM'14, Chicago, IL,
  USA, August 17-22, 2014}, pages 175--186, 2014.
\newblock \url{https://doi.org/10.1145/2619239.2626311}.

\bibitem{calcadan-osdi-2020}
Joshua Fried, Zhenyuan Ruan, Amy Ousterhout, and Adam Belay.
\newblock Caladan: Mitigating interference at microsecond timescales.
\newblock In {\em 14th {USENIX} Symposium on Operating Systems Design and
  Implementation, {OSDI} 2020, Virtual Event, November 4-6, 2020}, pages
  281--297, 2020.
\newblock \url{https://www.usenix.org/conference/osdi20/presentation/fried}.

\bibitem{los-apnet-2017}
Yukai Huang, Jinkun Geng, Du~Lin, Bin Wang, Junfeng Li, Ruilin Ling, and Dan
  Li.
\newblock {LOS:} {A} high performance and compatible user-level network
  operating system.
\newblock In {\em Proceedings of the First Asia-Pacific Workshop on Networking,
  APNet 2017, Hong Kong, China, August 3-4, 2017}, pages 50--56, 2017.
\newblock \url{https://doi.org/10.1145/3106989.3106997}.

\bibitem{WEB:SeaStar}
Seastar.
\newblock \url{https://github.com/scylladb/seastar}.

\bibitem{socksdirect-2019}
Bojie Li, Tianyi Cui, Zibo Wang, Wei Bai, and Lintao Zhang.
\newblock Socksdirect: datacenter sockets can be fast and compatible.
\newblock In {\em Proceedings of the {ACM} Special Interest Group on Data
  Communication, {SIGCOMM} 2019, Beijing, China, August 19-23, 2019}, pages
  90--103, 2019.

\bibitem{erpc-2019}
Anuj Kalia, Michael Kaminsky, and David Andersen.
\newblock Datacenter rpcs can be general and fast.
\newblock In {\em 16th {USENIX} Symposium on Networked Systems Design and
  Implementation, {NSDI} 2019, Boston, MA, February 26-28, 2019}, pages 1--16,
  2019.

\bibitem{snap-sosp-2019}
Michael Marty, Marc de~Kruijf, Jacob Adriaens, Christopher Alfeld, Sean Bauer,
  Carlo Contavalli, Michael Dalton, Nandita Dukkipati, William~C. Evans, Steve
  Gribble, Nicholas Kidd, Roman Kononov, Gautam Kumar, Carl Mauer, Emily
  Musick, Lena~E. Olson, Erik Rubow, Michael Ryan, Kevin Springborn, Paul
  Turner, Valas Valancius, Xi~Wang, and Amin Vahdat.
\newblock Snap: a microkernel approach to host networking.
\newblock In {\em Proceedings of the 27th {ACM} Symposium on Operating Systems
  Principles, {SOSP} 2019, Huntsville, ON, Canada, October 27-30, 2019}, pages
  399--413, 2019.

\bibitem{perfiso-atc18}
Calin Iorgulescu, Reza Azimi, Youngjin Kwon, Sameh Elnikety, Manoj Syamala,
  Vivek~R. Narasayya, Herodotos Herodotou, Paulo Tomita, Alex Chen, Jack Zhang,
  and Junhua Wang.
\newblock Perfiso: Performance isolation for commercial latency-sensitive
  services.
\newblock In {\em 2018 {USENIX} Annual Technical Conference, {USENIX} {ATC}
  2018, Boston, MA, USA, July 11-13, 2018}, pages 519--532, 2018.
\newblock
  \url{https://www.usenix.org/conference/atc18/presentation/iorgulescu}.

\bibitem{arachne-osdi-2018}
Henry Qin, Qian Li, Jacqueline Speiser, Peter Kraft, and John~K. Ousterhout.
\newblock Arachne: Core-aware thread management.
\newblock In {\em 13th {USENIX} Symposium on Operating Systems Design and
  Implementation, {OSDI} 2018, Carlsbad, CA, USA, October 8-10, 2018}, pages
  145--160, 2018.
\newblock \url{https://www.usenix.org/conference/osdi18/presentation/qin}.

\end{thebibliography}
